\documentclass[sort&compress]{elsarticle}
\usepackage{color}	
\usepackage{graphicx}   
\usepackage{amsmath}
\usepackage{amsfonts}
\usepackage{enumitem}
\usepackage{here}
\usepackage{listings}
\newcommand{\FSmall}{\scriptsize}
\usepackage[top=30truemm,bottom=30truemm,left=25truemm,right=25truemm]{geometry}

\newcommand{\Bold}[1]{{\boldsymbol #1}}

\newcommand{\Half}{\frac{1}{2}}
\newcommand{\sgn}{\operatorname{sgn}}
\newcommand{\exptw}{{\rm exp2}}
\newcommand{\logtw}{{\rm log2}}
\newcommand{\Til}{\widetilde}
\newcommand{\abs}[1]{\lvert#1\rvert}
\newcommand{\rA}{{\rm A}}
\newcommand{\rB}{{\rm B}}
\newcommand{\phiOp}{\phi^*}
\newcommand{\g}{g}
\newcommand{\absg}{\abs{g}}
\newcommand{\pred}{\operatorname{pred}}

\newcommand{\revisedBegin}{\color{black}}
\newcommand{\revisedEnd}{\color{black}}
\newcommand{\revised}[1]{\textcolor{black}{#1}}

\allowdisplaybreaks[1]
\journal{Computers \& Fluids} 
\begin{document}
\title{A simple volume-of-fluid reconstruction method
for three-dimensional two-phase flows}
\author{Akio Kawano\corref{cor1}}
\cortext[cor1]{Corresponding author. Tel.:+81 45 778 5871; fax: +81 45 778 5491.}
\ead{kawanoa@jamstec.go.jp}
\address{
Japan Agency for Marine-Earth Science and Technology,
3173-25, Showa-machi, Kanazawa-ku, Yokohama, 236-0001, Japan}
\begin{abstract}
A new PLIC (piecewise linear interface calculation)-type
VOF (volume of fluid) method, called APPLIC (approximated PLIC) method, 
is presented.
Although the PLIC method is one of the most accurate VOF methods,
the three-dimensional algorithm is complex.
Accordingly, it is hard to develop and maintain the computational code.
The APPLIC method reduces the complexity using simple approximation formulae.
Three numerical tests were performed to compare the accuracy of
the SVOF (simplified volume of fluid),
VOF/WLIC (weighed line interface calculation),
THINC/SW (tangent of hyperbola for interface capturing/slope weighting),
THINC/WLIC, PLIC, and APPLIC methods.
The results of the tests show that the APPLIC results are as accurate as the PLIC results
and are more accurate than the SVOF, VOF/WLIC, THINC/SW, and THINC/WLIC results.
It was demonstrated that
the APPLIC method is more computationally efficient than the PLIC method.
\end{abstract}
\maketitle

\section{Introduction}
Two-phase flows are essential in many research fields;
for example, in relation to cloud and precipitation droplets in the atmosphere,
water waves, cooling devices, oil and gas pipelines,
chemical industrial plants,
and thermal power stations.
In the recent decades, many interface tracking methods for simulating
two-phase flows have been developed.
The VOF (volume of fluid) method,
originated by Hirt and Nichols \cite{hirt1981volume},
is one of the most widely used algorithms.
Excellent reviews of the VOF method
have been given by Rudman \cite{rudman1997volume},
Rider and Kothe \cite{rider1998reconstructing},
Scardovelli and Zaleski \cite{scardovelli1999direct},
and Pilliod and Puckett \cite{pilliod2004second}.

The VOF method is based on
the spatial discretization of a characteristic function to distinguish
between two phases,
and the reconstruction of the interfaces for advection.
Suppose that we wish to simulate an incompressible two-phase (`light' and `dark') flow
in the three-dimensional Cartesian space $\Bold{x}=(x_1,x_2,x_3)$.
The characteristic function for the flow is defined as
\begin{align}
\chi(\Bold{x})=
\begin{cases}
0 &\text{if there is light fluid at point $\Bold{x}$},\\
1 &\text{if there is dark fluid at point $\Bold{x}$}.
\end{cases}
\end{align}
The interfaces between the two phases are represented by
the discontinuity of the characteristic function.
In this paper, we suppose that a computational grid composed of 
cubic cells of a edge $\Delta x$ is used.
Extension of our analysis to general regular grids is straightforward.
By descretizing the characteristic function in a computational cell $(i,j,k)$,
we can obtain the volume fraction
\begin{align}
C_{i,j,k} &=
\frac{1}{(\Delta x)^3}
\int_{\Omega_{i,j,k}}
\chi(\Bold{x})\,d\Bold{x},
\end{align}
where $\Omega_{i,j,k}$ is the domain of the cell.
It is obvious from the definition that
\begin{align}
C_{i,j,k}\begin{cases}
=0       &\text{if the cell is filled by light fluid},\\
\in(0,1) &\text{if the cell contains both fluids (interface cell)},\\
=1       &\text{if the cell is filled by dark fluid}.
\end{cases}
\end{align}

The VOF method reconstructs the shape of the interface
in each interface cell
to evaluate VOF advection fluxes.
Various schemes for VOF reconstruction have been presented.
The PLIC (piecewise linear interface calculation) method \cite{youngs1982,li1995}
reconstructs an interface in a cell as a plane (in three-dimensional space) or
a line (in two-dimensional space) with a given normal vector.
The SLIC (simple line interface calculation) method \cite{noh1976}
assumes the shape of an interface to be
a plane parallel to one of the cell faces.
The VOF/WLIC (weighted line interface calculation) method \cite{yokoi2007efficient}
evaluates an advection flux through a cell face as a weighted sum of SLIC fluxes.
The SVOF (simplified volume of fluid) method \cite{marek2008} is similar to
the VOF/WLIC method, except for the weight formula.
In the THINC (tangent of hyperbola for interface capturing) method \cite{xiao2005simple},
interfaces are represented by the use of the hyperbolic tangent.
Improved THINC methods have also been
proposed \cite{yokoi2007efficient,xiao2011revisit,ii2012interface}.

Although it is known to be one of the most accurate reconstruction methods,
a three-dimensional implementation of the PLIC method is a troublesome task.
The PLIC method requires the solution of two geometric problems,
as to a cut-volume of a cube by a plane,
which are very complicated
especially in three-dimensional cases.
Scardovelli and Zaleski have
provided two sophisticated algorithms (hereafter called the SZ algorithms)
to solve these problems \cite{scardovelli2000analytical}.
Although the SZ algorithms make the implementation of the PLIC method easier
because of their compactness,
these are still too complex for quick and easy implementation.
Computational routines that implement the SZ algorithms must involve multiple ``if'' statements,
which make it hard to develop and maintain the routines,
and potentially inhibit its optimal compilation,
especially for processors susceptible to conditional branches,
e.g., deeply pipelined processors, processors with SIMD (single
instruction multiple data) operations,
vector processors, and GPUs (graphics processing units) \cite{hennessy2006computer}.





In this paper, a PLIC-type VOF method
called the APPLIC (approximated PLIC) method is presented.
In the APPLIC method,
interfaces are reconstructed in a similar manner as in the PLIC method,
except that the geometric problems are solved through the use of simple approximation formulae.

This paper is organized as follows.
In Section 2, we describe the APPLIC method.
Section 3 compares the accuracy and computational efficiency of the APPLIC method
with some existing VOF methods.
Finally, conclusions are summarized in Section 4.

The following vector notation is used throughout this paper.
Bold letters denote three-dimensional vectors and
the corresponding non-bold letters with subscripts 1, 2, or 3 denote the
vector components.
For example, $\Bold{u}=(u_1,u_2,u_3)$ and
$\Bold{m}''_\rA = (m''_{\rA,1}, m''_{\rA,2}, m''_{\rA,3})$.
\revised{We use $\|\Bold{m}\|_n$ to represent the $n$-norm of $\Bold{m}$;
namely, $\|\Bold{m}\|_1 = \abs{m_1} + \abs{m_2} + \abs{m_3}$
and $\|\Bold{m}\|_2 = (m_1^2 + m_2^2 + m_3^2)^{1/2}$.}
The expression $\Bold{m}\ge a$ stands for the condition $m_1\ge a$, $m_2\ge a$, and $m_3\ge a$.

\section{Method}
\subsection{The PLIC method using the SZ algorithms}
In this paper,
we use directional splitting for advection and
a regular staggered grid 
where velocity components $u_1$, $u_2$, and $u_3$ are stored at the centers of the cell faces
$\{(i+1/2,j,k)\}$,
$\{(i,j+1/2,k)\}$, and
$\{(i,j,k+1/2)\}$, respectively.
We assume that the Courant-Friedrichs-Lewy (CFL) condition,
\begin{align}
\frac{\abs{u_l}\Delta t}{ \Delta x} < 1 \quad\text{for all $l\in\{1,2,3\}$},
\end{align}
holds, where $\Delta t$ is the time step size.

\begin{figure}[H]
\centering
\includegraphics[width=6cm]{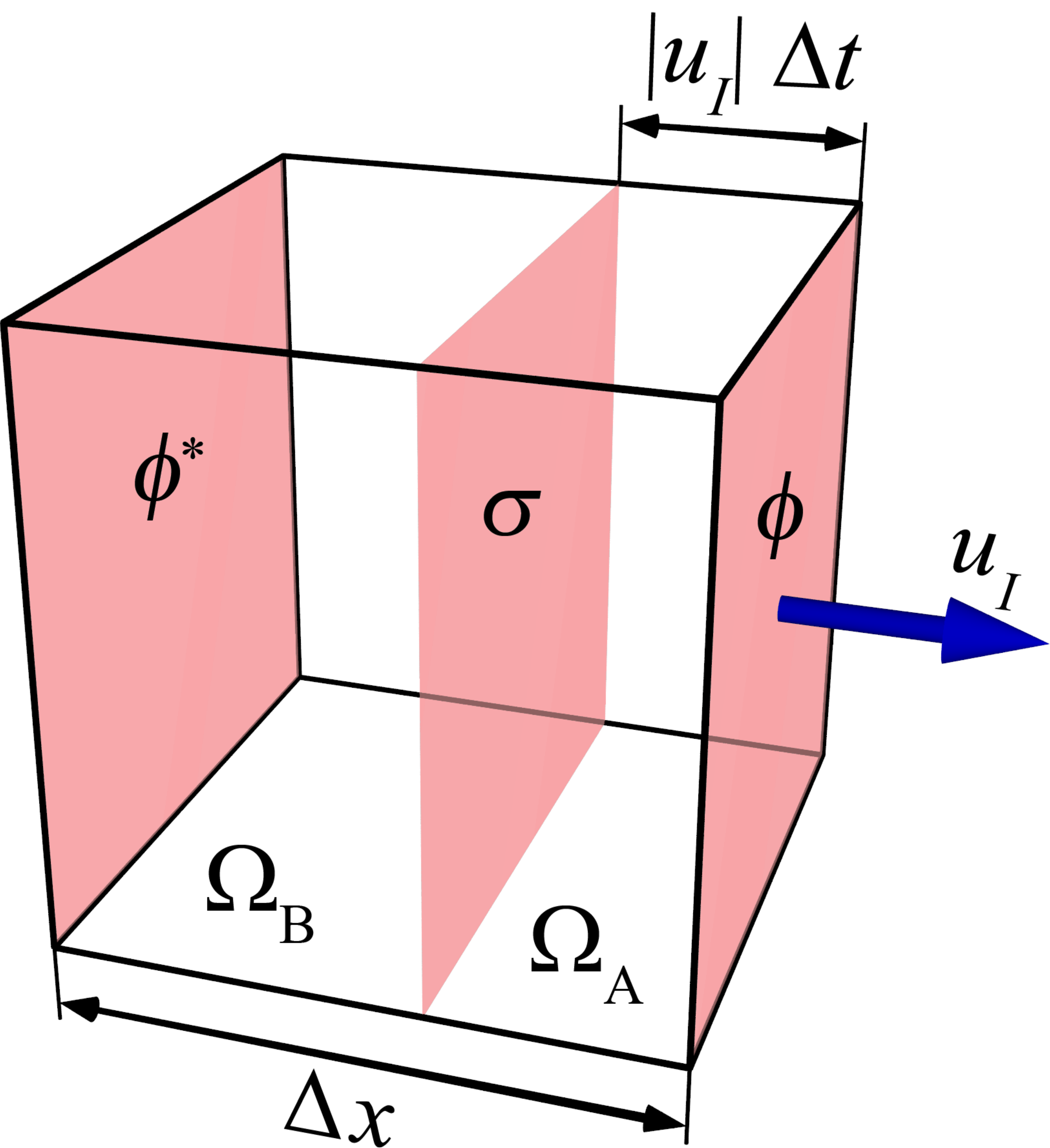}
\caption{Schematic illustration of a donor cell
with respect to a flux through a cell face $\phi$.
The cell is divided into two subcells,
$\Omega_{\rm A}$ and $\Omega_{\rm B}$,
by the section $\sigma$
parallel to $\phi$ and laid $\abs{u_I}\Delta t$ away from $\phi$.}
\label{fig_schematic_illust}
\end{figure}

Let
$\phi$ be the face, and
$u_I$ ($I=1$, 2, or 3)
the velocity component placed on $\phi$.
Let $\Omega$ be the donor cell,
which is the cell that has $\phi$ as a cell face and lies on the upwind side of $u_I$.
Let $\phiOp$ be the opposite face of the face $\phi$ in the cell $\Omega$.
The cell $\Omega$ is partitioned into two subcells by the section $\sigma$
parallel to $\phi$ and laid $\abs{u_I}\Delta t$ away from $\phi$.
Let $\Omega_\rA$ and $\Omega_\rB$
be the subcells of $\Omega$
between $\phi$ and $\sigma$
and between $\sigma$ and $\phiOp$,
respectively (see Fig.~\ref{fig_schematic_illust}).
The section $\sigma$ is always located between $\phi$ and $\phi^*$
because of the CFL condition.
Let $C_\rA$ and $C_\rB$ be
the partial volume fractions
in $\Omega_\rA$ and $\Omega_\rB$, respectively, defined as
\begin{align}
C_\rA &=
\frac{1}{(\Delta x)^3}
\int_{\Omega_\rA}
\chi(\Bold{x})\,d\Bold{x},\\
C_\rB &=
\frac{1}{(\Delta x)^3}
\int_{\Omega_\rB}
\chi(\Bold{x})\,d\Bold{x}.
\end{align}
It is obvious that $C_\rA\in[0,1]$ and $C_\rB\in[0,1]$.
From the definitions, we have
\begin{align}
C_\rA + C_\rB&= C,\label{eq:c_eq_cacb}
\end{align}
where $C$ is the volume fraction in the donor cell $\Omega$.

The computational advection flux of the volume fraction through the face $\phi$
(i.e., the amount of the volume fraction through the face during $\Delta t$)
is obtained via
\begin{align}
F=C_\rA\sgn u_I,\label{eq:f_phi_eq_c_a_sign_u_I}
\end{align}
where $\sgn$ is the sign function defined as
\begin{align}
\sgn x&=
\begin{cases}
 1 &\text{if $x \ge 0$},\\
-1 &\text{if $x < 0$}.
\end{cases}
\end{align}
In some cases, $C_\rA$ is easily determined
by
\begin{align}
C_\rA =
\begin{cases}
0 &\text{if $C=0$ or $\absg=0$},\\
\absg &\text{if $C=1$},
\end{cases}
\end{align}
where $\absg$ denote the local CFL number in the cell $\Omega$
with respect to the flux through the face $\phi$:
\begin{align}
\g=\frac{u_I\Delta t}{\Delta x}.
\end{align}
Because of the CFL condition, $\g$ must be in the range $(-1,1)$.
If $C\in (0,1)$ and $\absg > 0$,
$C_\rA$ is determined through the reconstruction
of the interface in the donor cell $\Omega$.

Here, we define the two geometric problems crucial to the PLIC method,
which are mutually inverse.
Consider a unit cube
$
U=\{\Bold{x}\in[0,1]^3\}
$
and an oriented plane
$
P(\alpha, \Bold{m})=
\{\Bold{x} |\, \Bold{m}\cdot\Bold{x} < \alpha\},
$
where $\Bold{m}$ is the normal vector of the plane,
and $\alpha$ is the plane constant.
Note that an oriented plane is not a thin object without volume,
but is a solid object with an inside and an outside.
Let $V$ be the volume of the intersection
between the unit cube and the oriented plane.
One of the problems, called the forward problem,
is to determine the value of $V$ for given
$\alpha$ and $\Bold{m}$.
The other problem, called the inverse problem, is
to determine the value of $\alpha$ for given $V$ and $\Bold{m}$.
Namely,
\begin{align}
V(\alpha, \Bold{m})
&=
\int_{U\cap P(\alpha, \Bold{m})}
d\Bold{x},\\
\alpha(V',\Bold{m})
&=
\alpha'\quad\text{such that\quad
$V(\alpha', \Bold{m})=V'$.}
\end{align}
To reduce the complexity of the problems,
the SZ algorithms restrict $\Bold{m}$ to a vector
so that $\Bold{m}\ge0$ and $\|\Bold{m}\|_1=1$.

%

The functions $V(\alpha,\Bold{m})$ and $\alpha(V,\Bold{m})$
have the following properties~\cite{scardovelli1999direct}.
\begin{enumerate}[label=(\Roman*),ref=\Roman*]
\item\label{prop0}
The value of $V(\alpha,\Bold{m})$ is within the range $[0,1]$
and
\begin{align}
V(\alpha,\Bold{m})=
\begin{cases}
0 & \text{if $\alpha \le 0$},\\
1 & \text{if $\alpha \ge 1$}.\\
\end{cases}\label{eq_V_outof_range}
\end{align}
\item\label{prop1}
The value of $\alpha(V,\Bold{m})$ is within the range $[0,1]$ for $V\in[0,1]$.
\item\label{propA}
The functions $V$ and $\alpha$ are invariant with respect to
a permutation of $m_1$, $m_2$, and $m_3$.
\item\label{propC}
The functions $V$ and $\alpha$ are continuous, one-to-one, and monotonically
increasing functions of
$\alpha$ and $V$ in the ranges $V\in[0,1]$ and $\alpha\in[0,1]$, respectively.
\item\label{propD}
The first derivatives $\partial V/ \partial \alpha$ and
$\partial \alpha/\partial V$ are continuous
and monotonically nondecreasing
functions of $\alpha\in(0, 1/2]$ and $V\in(0,1/2]$, respectively.
\item\label{propE}
The curve $(\alpha,V)$ passes through the points
$(0,0)$, $(1/2,1/2)$, and $(1,1)$.
\item\label{propG}
If $\Bold{m}=(0,0,1)$ and $(0,1/2,1/2)$,
the values of $V(\alpha,\Bold{m})$ become
$\alpha$ and $(2\alpha)^2/2$, respectively,
for $\alpha\in[0,1/2]$.
\item
\label{propH}
The curve $(\alpha,V)$ has point symmetry
(or odd symmetry) with respect to
$(\alpha,V) = (1/2, 1/2)$;
namely,
\begin{align}
V(\alpha,\Bold{m})&=
1-V(1-\alpha,\Bold{m}),\label{symmetry_V}\\
\alpha(V,\Bold{m})&=
1-\alpha(1-V,\Bold{m}).\label{symmetry_alpha}
\end{align}
\end{enumerate}
All the properties except \eqref{propG} hold for the arbitrary $\Bold{m}$.

Figure \ref{fig.szcode} shows an example implementation of
the SZ algorithms written in Fortran 90.
The cbrt function, used in line 44,
is an intrinsic function that returns the real cube root of the argument.
Although this is not included in the Fortran 90 standard,
many Fortran compilers support this.
The abs functions in lines \revised{11} and \revised{40} are used to prevent vm2 being negative
owing to the numerical error in floating point arithmetic when $\text{vm3} \simeq 1$.
\begin{figure}[H]
\renewcommand*\thelstnumber{\roman{lstnumber}}
\vspace*{-3em}
\begin{lstlisting}
module constants
  real(8), parameter :: CONST_TINY = 1d-25          ! an extremely small constant
  real(8), parameter :: CONST_PI = 3.14159265358979323846d0 ! $\pi$
end module constants
\end{lstlisting}
\vspace*{-0.5em}
\renewcommand*\thelstnumber{\arabic{lstnumber}}
\begin{lstlisting}
  function calc_v(alpha, vma, vmb, vmc) result(v)
    ! Preconditions:$\; {\tt vma}\in[0,1],\;{\tt vmb}\in[0,1],\;{\tt vmc}\in[0,1],\;%
{\tt vma} + {\tt vmb} + {\tt vmc} = 1$.
    use constants
    real(8), intent(in) :: alpha, vma, vmb, vmc
    real(8) :: v, a, vm1, vm2, vm3, vm12
    a = min(alpha, 1d0 - alpha)
    v = 0d0
    if (a > 0d0) then
      vm1 = min(vma, vmb, vmc)  $\qquad\,$! Sort $\,{\tt vm1}, {\tt vm2}, {\tt vm3}$
      vm3 = max(vma, vmb, vmc)  $\qquad$! such that $\,{\tt vm1} \le {\tt vm2} \le {\tt vm3}\,$.
      vm2 = abs(1d0 - vm3 - vm1)
      vm12 = vm1 + vm2
      if (a < vm1) then
        v = a ** 3 / (6d0 * vm1 * vm2 * vm3)
      else if (a < vm2) then
        v = a * (a - vm1) / (2d0 * vm2 * vm3) + &
            vm1 ** 2 / (6d0 * vm2 * vm3 + CONST_TINY)
      else if (a < min(vm12, vm3)) then
        v = (a ** 2 * (3d0 * vm12 - a) + vm1 ** 2 * (vm1 - 3d0 * a) + &
             vm2 ** 2 * (vm2 - 3d0 * a)) / (6d0 * vm1 * vm2 * vm3)
      else if (vm3 < vm12) then
        v = (a ** 2 * (3d0 - 2d0 * a) + vm1 ** 2 * (vm1 - 3d0 * a) +    &
             vm2 ** 2 * (vm2 - 3d0 * a) + vm3 ** 2 * (vm3 - 3d0 * a)) / &
             (6d0 * vm1 * vm2 * vm3)
      else
        v = (a - 0.5d0 * vm12) / vm3
      end if
    end if
    if (alpha > 0.5d0) v = 1d0 - v
  end function calc_v

  function calc_alpha(v, vma, vmb, vmc) result(alpha)
    ! Preconditions:$\; {\tt v}\in(0,1),%
        \; {\tt vma}\in[0,1],\;{\tt vmb}\in[0,1],\;{\tt vmc}\in[0,1],\;{\tt vma} + {\tt vmb} + {\tt vmc} = 1$.
    use constants
    real(8), intent(in) :: v, vma, vmb, vmc
    real(8) :: alpha, w, vm1, vm2, vm3, vm12, v1, v3, a0, a1, a2, q0, sp, th
    w = min(v, 1d0 - v)
    vm1 = min(vma, vmb, vmc)  $\qquad\,$! Sort $\,{\tt vm1}, {\tt vm2}, {\tt vm3}$
    vm3 = max(vma, vmb, vmc)  $\qquad$! such that $\,{\tt vm1} \le {\tt vm2} \le {\tt vm3}\,$.
    vm2 = abs(1d0 - vm3 - vm1)
    vm12 = vm1 + vm2
    v1 = vm1 ** 2 / (6d0 * vm2 * vm3 + CONST_TINY)
    if (w < v1) then
      alpha = cbrt(6d0 * vm1 * vm2 * vm3 * w)
    else if (w < v1 + (vm2 - vm1) / (2.0 * vm3)) then
      alpha = 0.5d0 * (vm1 + sqrt(vm1 ** 2 + 8d0 * vm2 * vm3 * (w - v1)))
    else
      alpha = 0d0
      if (vm3 < vm12) then
        v3 = (vm3 ** 2 * (3d0 * vm12 - vm3) + vm1 ** 2 * (vm1 - 3d0 * vm3) + &
              vm2 ** 2 * (vm2 - 3d0 * vm3)) / (6d0 * vm1 * vm2 * vm3)
      else
        v3 = 0.5d0 * vm12 / vm3
        if (v3 <= w) alpha = vm3 * w + 0.5d0 * vm12
      end if
      if (alpha == 0.0) then
        if (w < v3) then
          a2 = -3d0 * vm12
          a1 = 3d0 * (vm1 ** 2 + vm2 ** 2)
          a0 = -(vm1 ** 3 + vm2 ** 3 - 6d0 * vm1 * vm2 * vm3 * w)
        else
          a2 = -1.5d0
          a1 = 1.5d0 * (vm1 ** 2 + vm2 ** 2 + vm3 ** 2)
          a0 = -0.5d0 * (vm1 ** 3 + vm2 ** 3 + vm3 ** 3 - 6d0 * vm1 * vm2 * vm3 * w)
        end if
        q0 = (1d0/6d0) * (a1 * a2 - 3d0 * a0) - a2 ** 3 * (1d0/27d0)
        sp = sqrt(-1d0/3d0 * a1 + 1d0/9d0 * a2 ** 2)
        th = 1d0/3d0 * acos(q0 / (sp ** 3))
        alpha = 2d0 * sp * cos(th + (4d0/3d0 * CONST_PI)) - (1d0/3d0) * a2
      end if
    end if
    if (v > 0.5d0) alpha = 1d0 - alpha
  end function calc_alpha
\end{lstlisting}
\caption{Example implementation of
the SZ algorithms written in Fortran 90.}
\label{fig.szcode}
\def\thepage{}
\end{figure}
\clearpage

\revised{By using} the SZ algorithms
to evaluate the functions $V(\alpha, \Bold{m})$ and $\alpha(V, \Bold{m})$,
the volume fraction $C_\rA$ for the PLIC method is determined
as
\begin{subequations}
\begin{align}
C_\rA(g, C, \Bold{n}) = \absg\, V(\alpha''_\rA, \Bold{m}''_\rA),
\end{align}
with
\label{eq_A_set}
\begin{align}
\alpha''_\rA&=
\begin{cases}
Q'_\rA \alpha'&           \text{if $n_I\, \g  \le 0$},\\
Q'_\rA (\alpha' - r_\rA)& \text{if $n_I\, \g > 0$},
\end{cases}
\label{eq_A_app}\\
m''_{\rA,l} &=
\begin{cases}
Q'_\rA m'_l \absg & \text{if $l=I$},\\
Q'_\rA m'_l & \text{if $l\ne I$},
\end{cases}
\\
Q'_\rA&=\frac{1}{1-r_\rA},\\
r_\rA &= m'_I(1-\absg),\\
\alpha'&=\alpha(C,\Bold{m}'),\label{eq_A_ap}\\
m'_l&=\frac{\abs{n_l}}{\|\Bold{n}\|_1},\label{eq_A_ml}
\end{align}
\end{subequations}
where $l$ is an index running from one to three, and
$\Bold{n}$ is
the normal vector of the interface in the donor cell $\Omega$
oriented from the dark fluid to the light fluid.
See Appendix A for the derivation of the equations.
The vector $\Bold{n}$ is 
determined as
\begin{align}
\Bold n=-\nabla C,
\end{align}
where $\nabla C$ is a numerical gradient of the volume fraction.
Accurate evaluation of numerical gradients is required
for accurate results.
Various algorithms
to evaluate numerical gradients for the VOF method
can be found
in the literature~\cite{%
pilliod2004second,%
scardovelli2003interface,%
aulisa2007interface,%
weymouth2010conservative,%
Wu2013739,%
vignesh2013noniterative}.

Figure \ref{fig.plic} shows an example implementation of
the PLIC method written in Fortran 90.
Here, the argument vn1 corresponds to $n_I$, and
vn2 and vn3 correspond to the other components of $\Bold{n}$.
\begin{figure}[H]
\begin{lstlisting}
  function calc_flux_plic(g, c, vn1, vn2, vn3) result(f)
    ! Preconditions:$\;{\tt g}\in(-1,1),\;{\tt c}\in(0,1).$
    use constants
    real(8), intent(in) :: g, c, vn1, vn2, vn3
    real(8) :: f, absg, alpha, qa, ra, vm1, vm2, vm3
    absg = abs(g)
    vm1 = abs(vn1)
    vm2 = abs(vn2)
    vm3 = abs(vn3) + CONST_TINY
    qa = 1d0 / (vm1 + vm2 + vm3)
    vm1 = vm1 * qa
    vm2 = vm2 * qa
    vm3 = vm3 * qa
    alpha = calc_alpha(c, vm1, vm2, vm3)
    ra = vm1 * (1d0 - absg)
    qa = 1d0 / (1d0 - ra)
    if (g * vn1 > ZERO) alpha = alpha - ra
    vm1 = vm1 * absg
    f = calc_v(alpha * qa, vm1 * qa, vm2 * qa, vm3 * qa) * g
  end function calc_flux_plic
\end{lstlisting}
\caption{Example implementation of
the PLIC algorithm written in Fortran 90.}
\label{fig.plic}
\end{figure}

\subsection{Approximation of the forward and the inverse problems}
A basic idea of the APPLIC method is to evaluate $V(\alpha,\Bold{m})$ and $\alpha(V,\Bold{m})$
by use of simple approximation formulae.
%
%
%
%
In the APPLIC method,
the function
$V(\alpha,\Bold{m})$ for $\alpha\in[0,1/2]$
is approximated
as
\begin{align}
\Til{V}(\alpha,\Bold{m}) &=
\frac{1}{2}(2\alpha)^{p},
\label{vf_alpha}
\end{align}
where $p$ is a positive-valued function of $\Bold{m}$.
The approximation of the
function $\alpha(V,\Bold{m})$ for $V\in[0,1/2]$
is derived by solving Eq.~\eqref{vf_alpha}
for $\alpha$:
\begin{align}
\Til{\alpha}(V,\Bold{m}) &=
\frac{1}{2}(2 V)^{1/p}.\label{alpha_vf}
\end{align}
Here a tilde ($\;\Til{\mbox{}}\;$) indicates an approximation.
Using Eqs.~\eqref{symmetry_alpha} and \eqref{alpha_vf},
we can give the formula of $\Til\alpha$ for $V\in[0,1]$ as
\begin{align}
\Til\alpha(V,\Bold{m}) &= 
\begin{cases}
\dfrac{1}{2}(2V)^{p}&\text{if $V \le 1/2$},\\
1 - \dfrac{1}{2}[2(1-V)]^{p}&\text{if $V > 1/2$}.
\end{cases}
\label{alpha_vf_2}
\end{align}
Similarly, using Eqs.~\eqref{eq_V_outof_range}, \eqref{symmetry_V}, and \eqref{vf_alpha},
we have
\begin{align}
\Til{V}(\alpha,\Bold{m}) &= 
\begin{cases}
0&\text{if $\alpha \le 0$},\\
\dfrac{1}{2}(2\alpha)^{p}&\text{if $0 < \alpha \le 1/2$},\\
1 - \dfrac{1}{2}[2(1-\alpha)]^{p}&\text{if $1/2 < \alpha < 1$},\\
1&\text{if $\alpha \ge 1$}.
\end{cases}
\label{vf_alpha_2}
\end{align}
The functions $\Til V$ and $\Til \alpha$ satisfy
properties \eqref{prop0}, \eqref{prop1}, \eqref{propC}, \eqref{propD}, \eqref{propE},
and \eqref{propH}.

\begin{figure}[H]
\centering
\includegraphics[width=11cm]{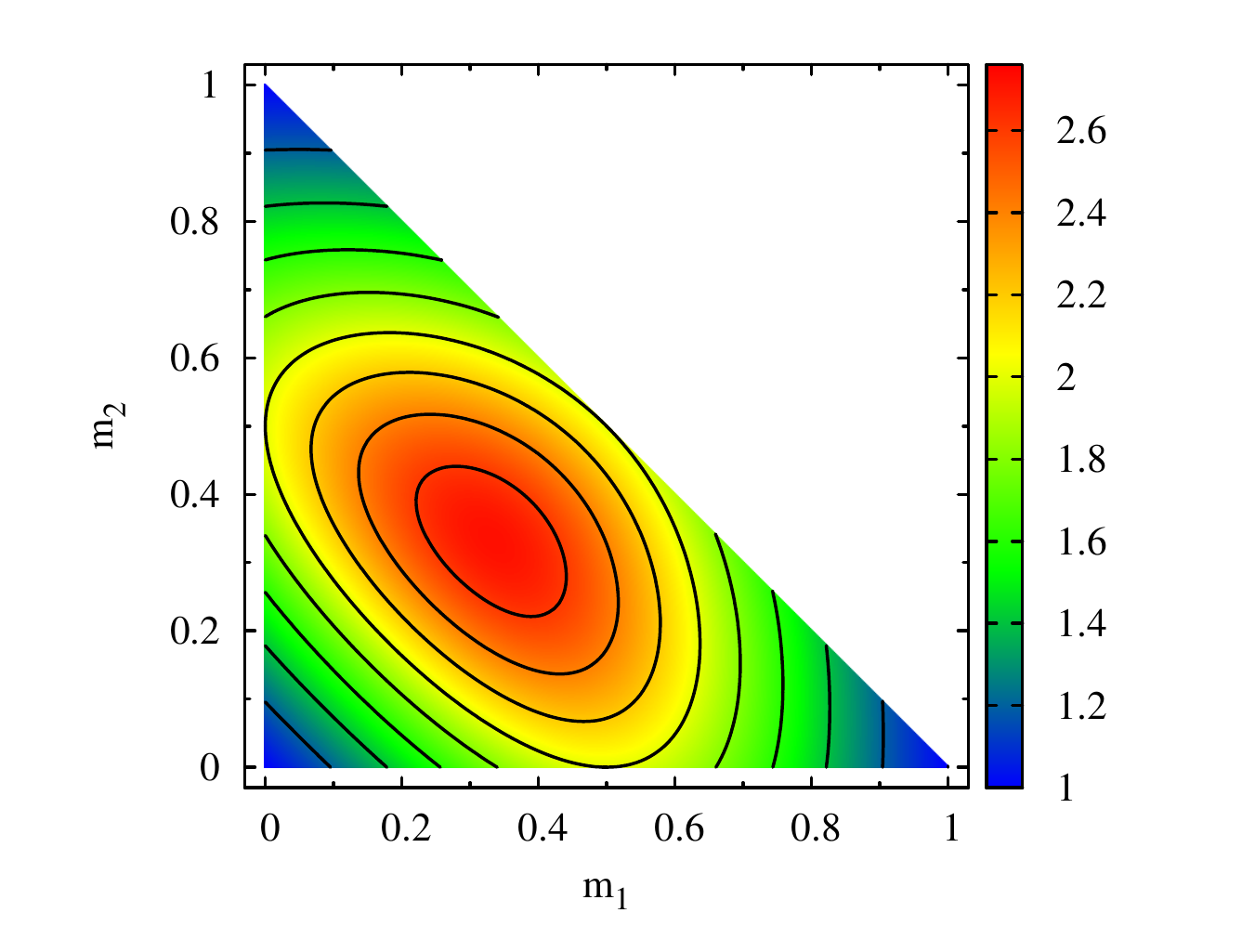}
\caption{Plot for $p_{\rm opt}$ as a function
of $\Bold{m}$. 
The component $m_3$ is given by $1 - m_1-m_2$.
The contour lines are drawn at $p_{\rm opt}=1$, 1.2, 1.4, $\ldots$, 2.6.}
\label{fig_p_opt}
\end{figure}

Let us formulate the function $p$.
%
First, we determine the optimal $p$, denoted by $p_{\rm opt}$,
that minimizes the square error $D$, defined as
\begin{align}
D\revised{(\Bold{m})} =\int_0^{\frac{1}{2}}[\Til V(\alpha,\Bold{m}) - V(\alpha,\Bold{m})]^2 d\alpha
+\int_0^{\frac{1}{2}}[\Til\alpha(V,\Bold{m}) - \alpha(V,\Bold{m})]^2 dV.
\label{error_d}
\end{align}
The function $p_{\rm opt}$ is plotted in Fig.~\ref{fig_p_opt}.

\begin{figure}[H]
\centering
\includegraphics[width=11cm]{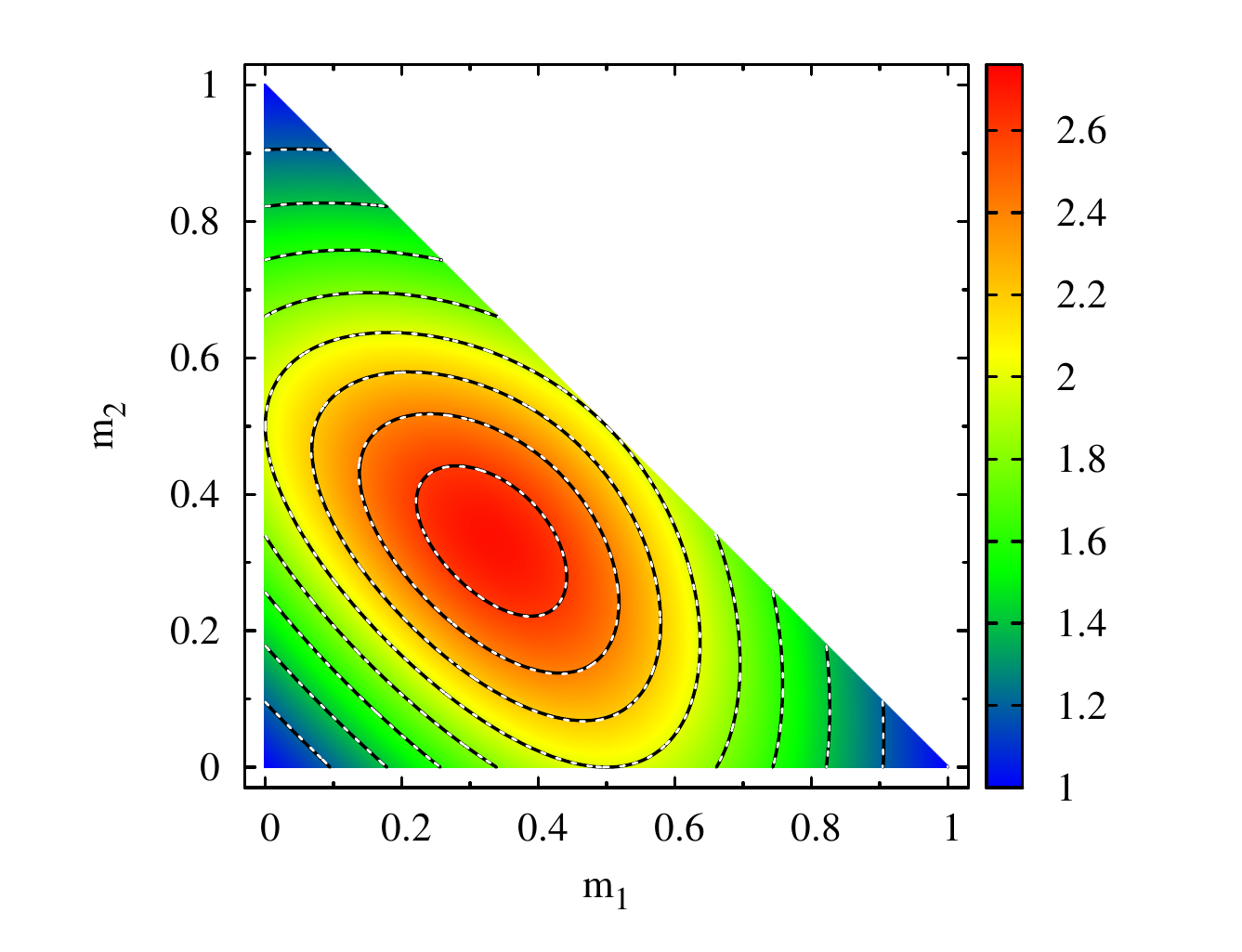}
\caption{Plot for $p$ (the color image and the black solid contour lines),
as obtained by Eq.~\eqref{approx_p_1} as a function
of $\Bold{m}$.
The component $m_3$ is given by $1 - m_1-m_2$.
The contour lines for $p_{\rm opt}$ are also shown as white dotted lines for comparison.
The contour lines are drawn at $p=1$, 1.2, 1.4, $\ldots$, 2.6.}
\label{fig_p}
\end{figure}

Next, we construct an arithmetic expression of $p$ that approximates $p_{\rm opt}$.
We found that $p_{\rm opt}$ can be approximated as
\begin{align}
p &= \frac{c_2 \xi^2+c_1 \xi + c_0}{\xi+c_0},\label{approx_p_1}
\end{align}
with
\begin{align}
\xi &= (b - m_1) (b - m_2) (b - m_3) - a,\label{eq_xi}
\end{align}
where $c_0$, $c_1$, $c_2$, $a$, and $b$ are constants.
To satisfy property \eqref{propA},
Eq.~\eqref{eq_xi} is designed to be symmetric with respect to
$m_1$, $m_2$, and $m_3$.
We \revised{impose} the following relations \revised{on the constants}
to satisfy property \eqref{propG}:
\begin{align}
c_0 &= \frac{b}{16}(c_2b + 4c_1 - 8),\\
a &= b^2(b-1).\label{eq_rel_a_b}
\end{align}
\revisedBegin
The optimal values of $b$, $c1$, and $c2$, shown in Table \ref{tbl_constants},
were determined by a least-square procedure
that minimizes the mean square error $\bar{D}$
with respect to $\Bold{m}$, defined as
\begin{align}
\bar{D}=\frac{\displaystyle \iint_{S_1} D\left(\frac{\Bold{n}}{\|\Bold{n}\|_1}\right) dS}
{\displaystyle \iint_{S_1}dS},
\end{align}
where $S_1$ is the part of the unit spherical surface in the first octant, namely,
\begin{align}
S_1 &= \{\Bold{x} \,|\, \text{$\Bold{x}\ge 0$ and $\|\Bold{x}\|_2 = 1$}\},
\end{align}
$\Bold{n}$ is a vector that scans $S_1$,
and $dS$ is the surface element.
\revisedEnd
Figure \ref{fig_p} shows $p$ given by Eq.~\eqref{approx_p_1}
with the optimal $b$, $c_1$, and $c_2$.
Here, $p_{\rm opt}$ is also plotted for comparison,
denoted by white and dotted contour lines.
The function $p$ given by Eq.~\eqref{approx_p_1}
fits with $p_{\rm opt}$ extremely well.
\begin{table}[H]
\caption{
Optimal values of the constants}
\centering
\begin{tabular}{cl}
\hline
$b$ & \revised{1.49} \\
$c_1$ & \revised{0.132}\\
$c_2$ & \revised{0.239} \\
\hline
\end{tabular}
\label{tbl_constants}
\end{table}

\begin{figure}[H]
\centering
\includegraphics[width=11cm]{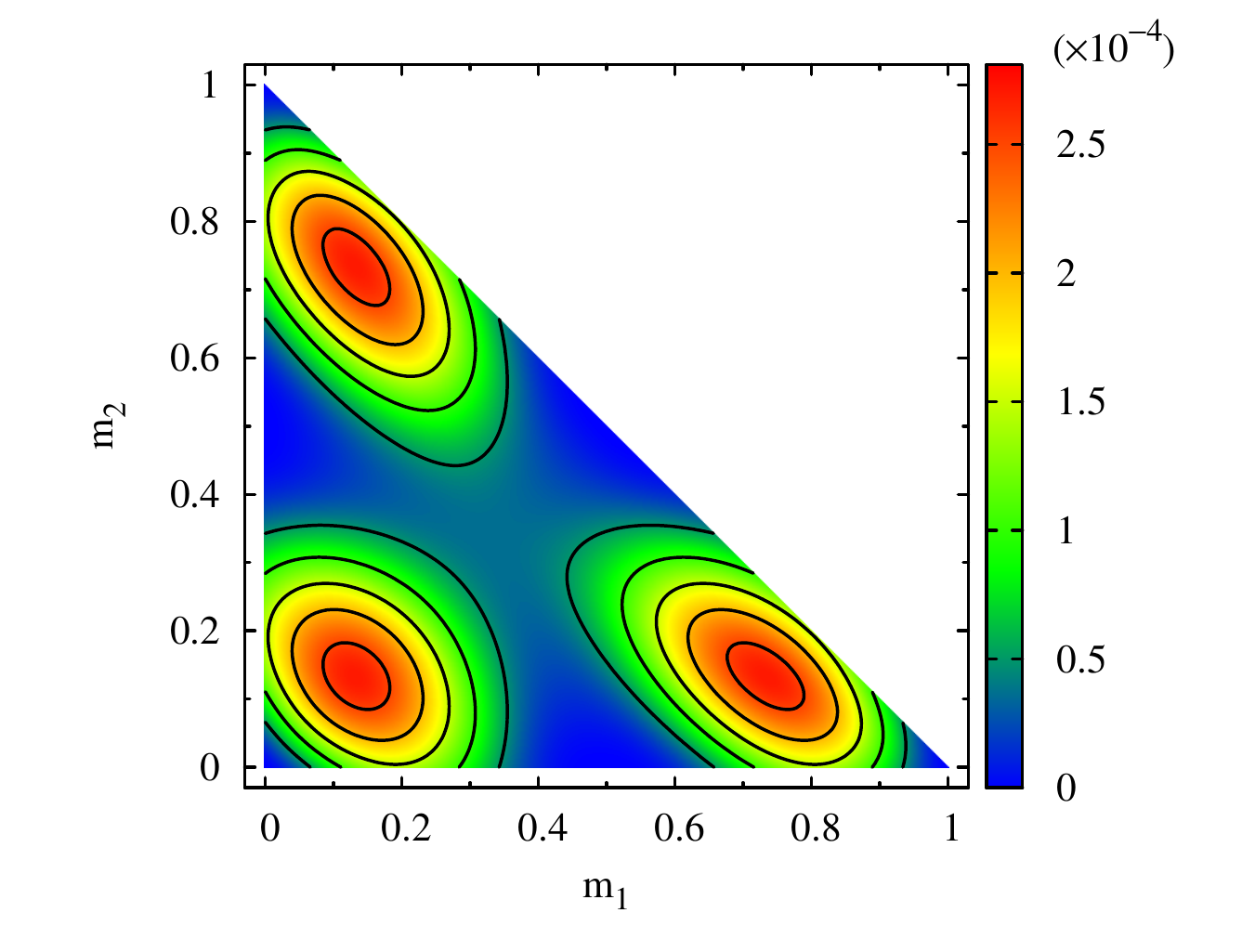}
\caption{Plot for the square error defined as Eq.~\eqref{error_d} as a function
of $\Bold{m}$.
The component $m_3$ is given by $1 - m_1-m_2$.
The contour lines are drawn at
$D=0.5\times 10^{-4}$, $1.0\times10^{-4}$,  $1.5\times10^{-4}$, $\ldots$,
$2.5\times10^{-4}$.}
\label{fig_error_d}
\end{figure}

\begin{figure}[H]
\centering
\includegraphics[width=13.0cm]{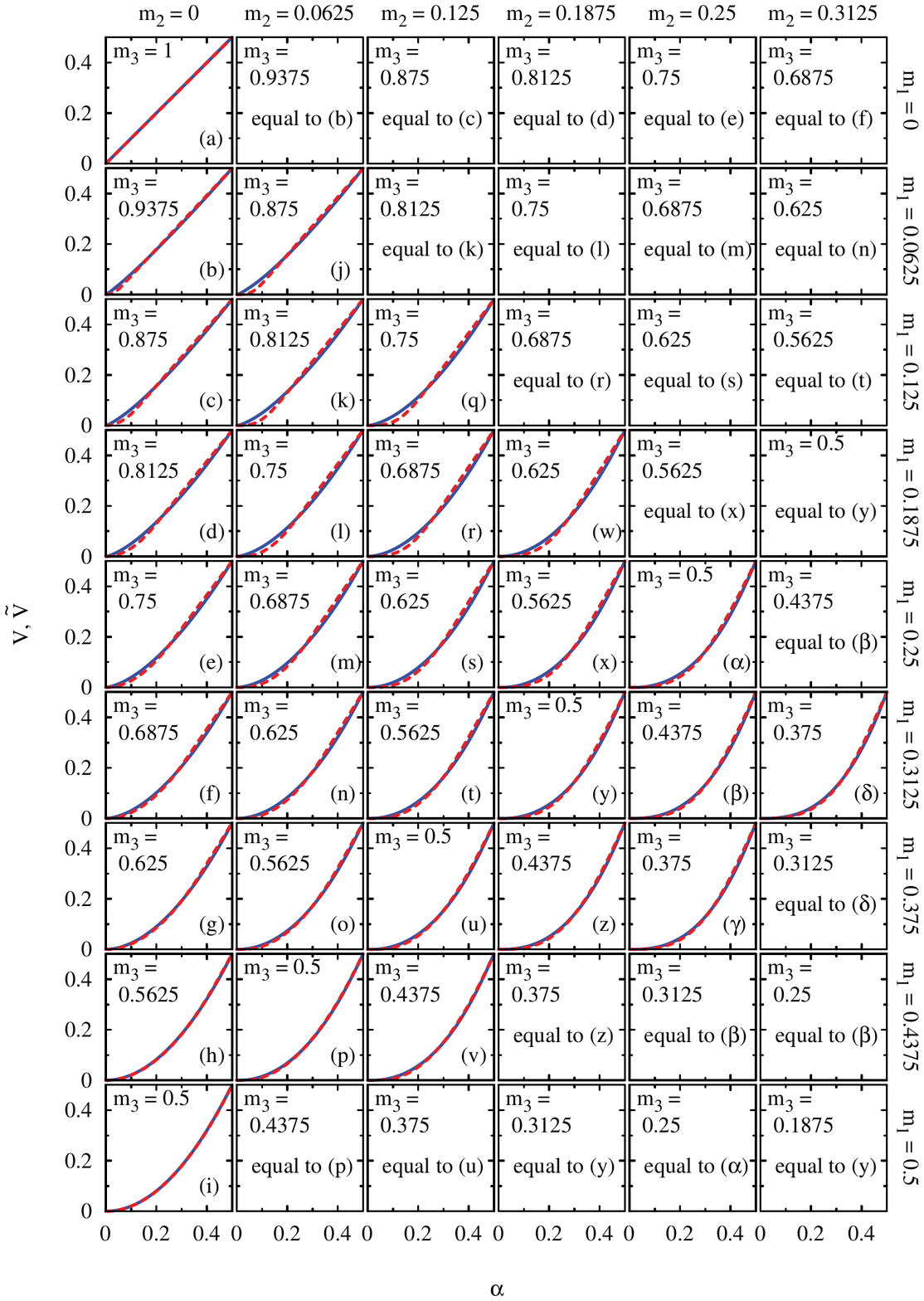}
\caption{Comparison between the functions $V(\alpha, \Bold{m})$ (\revised{dashed lines}) and
$\Til V(\alpha, \Bold{m})$ (\revised{solid lines})
as functions of $\alpha$ for various $\Bold{m}$.}
\label{fig_v_mat}
\end{figure}

The square error $D$
for $p$ obtained by Eq.~\eqref{approx_p_1}
is plotted in Fig.~\ref{fig_error_d}.
This figure shows that $D$ becomes the maximum, $D_{\rm max}=2.70\times10^{-4}$,
at $\Bold{m}=(0.734,0.133,0.133)$,
$(0.133,0.734,0.133)$, and 
$(0.133,0.133,0.734)$,
and becomes zero at
 $\Bold{m}=(1,0,0)$, $(0,1,0)$, $(0,0,1)$,
$(0,1/2,1/2)$, $(1/2,0,1/2)$, and $(1/2,1/2,0)$.
In Fig.~\ref{fig_v_mat}, we show $V$ and $\Til V$
as functions of $\alpha$
for various orientations of $\Bold{m}$.
In each pane,
curves $V(\alpha,\Bold{m})$ and $\Til V(\alpha,\Bold{m})$ are plotted for a specific $\Bold{m}$.
This figure shows that the discrepancy between
$V$ and $\Til V$ is sufficiently small for any $\Bold{m}$.
Panel (q) indicates the case of $\Bold{m}=(0.125,0.125,0,75)$,
which has the largest square error ($D=2.68\times10^{-4}$)
among the cases plotted in Fig.~\ref{fig_v_mat}.
Note that $(0.125,0.125,0,75)$ is close to $(0.133,0.133,0.734)$,
a case with the maximum square error.
\clearpage

Figure \ref{fig.va_approx} shows an example implementation of
the functions $\Til{V}$ and $\Til{\alpha}$ written in Fortran 90.
In lines \revised{11} and \revised{24}, the expression $\exp(\log(A) * B)$ is used instead of
the expression $A\, \mbox{$*\!*$}\, B$.
Both the expressions are mathematically equivalent if both $A$ and $B$ are positive.
Note that the variables {\tt a}, \revised{{\tt w},} {\tt p}, and {\tt invp} in Fig.~\ref{fig.va_approx}
are always positive.
However, the former expression requires slightly less computational cost
then the latter expression in most computer environments
because
the evaluation of $A\, \mbox{$*\!*$}\, B$
internally involves multiple conditional branches,
e.g., if $A$ is positive\revised{/zero}/negative, if $B$ is positive\revised{/zero}/negative,
and if $B$ is \revised{an integer or not}.
If C/C++ is used, the expression $\exptw(\logtw(A) * B)$ may
slightly more efficient.
\begin{figure}[H]
\renewcommand*\thelstnumber{\roman{lstnumber}}
\begin{lstlisting}
module applic_constants
  real(8), parameter :: PB  = 1.49d0, PC2 = 0.239d0, PC1 = 0.132d0, &
    PC0 = (PB * (PB * PC2 + 4d0 * PC1 - 8d0) / 16d0), &
    PA  = (PB * PB * (PB - 1d0))
end module applic_constants
\end{lstlisting}
\vspace*{-0.5em}
\renewcommand*\thelstnumber{\arabic{lstnumber}}
\begin{lstlisting}
  function calc_approx_v(alpha, vma, vmb, vmc) result(v)
    ! Preconditions:$\; {\tt vma}\in[0,1],\;{\tt vmb}\in[0,1],{\tt vmc}\in[0,1],\;{\tt vma} + {\tt vmb} + {\tt vmc} = 1$.
    use applic_constants
    real(8), intent(in) :: alpha, vma, vmb, vmc
    real(8) :: v, a, xi, p
    a = min(alpha, 1d0 - alpha)
    v = 0d0
    if (a > 0d0) then
        xi = (PB - vma) * (PB - vmb) * (PB - vmc) - PA
        p = ((PC2 * xi + PC1) * xi + PC0) / (xi + PC0)
        v = 0.5d0 * exp(log(a + a) * p)
    end if
    if (alpha > 0.5d0) v = 1d0 - v
  end function calc_approx_v

  function calc_approx_alpha(v, vma, vmb, vmc) result(alpha)
    ! Preconditions:$\; {\tt v}\in(0,1),%
        \; {\tt vma}\in[0,1],\;{\tt vmb}\in[0,1],\;{\tt vmc}\in[0,1],\;{\tt vma} + {\tt vmb} + {\tt vmc} = 1$.
    use applic_constants
    real(8), intent(in) :: v, vma, vmb, vmc
    real(8) :: alpha, w, xi, invp
    w = min(v, ONE - v)
    xi = (PB - vma) * (PB - vmb) * (PB - vmc) - PA
    invp = (xi + PC0) / ((PC2 * xi + PC1) * xi + PC0)
    alpha = 0.5d0 * exp(log(w + w) * invp)
    if (v > 0.5d0) alpha = 1d0 - alpha
  end function calc_approx_alpha
\end{lstlisting}
\caption{Example implementation of
the approximation functions $\Til{V}$ and $\Til{\alpha}$ written in Fortran 90.}
\label{fig.va_approx}
\end{figure}

\subsection{The crude APPLIC method}
Using the approximation functions $\Til{V}(\alpha,\Bold{m})$ and $\Til{\alpha}(V,\Bold{m})$
instead of functions ${V}(\alpha,\Bold{m})$ and ${\alpha}(V,\Bold{m})$
in Eq.~\eqref{eq_A_set},
we can determine computational advection fluxes via Eq.~\eqref{eq:f_phi_eq_c_a_sign_u_I}.
This straightforward method is called the crude APPLIC method.

\begin{figure}[H]
\centering
\includegraphics[width=8cm]{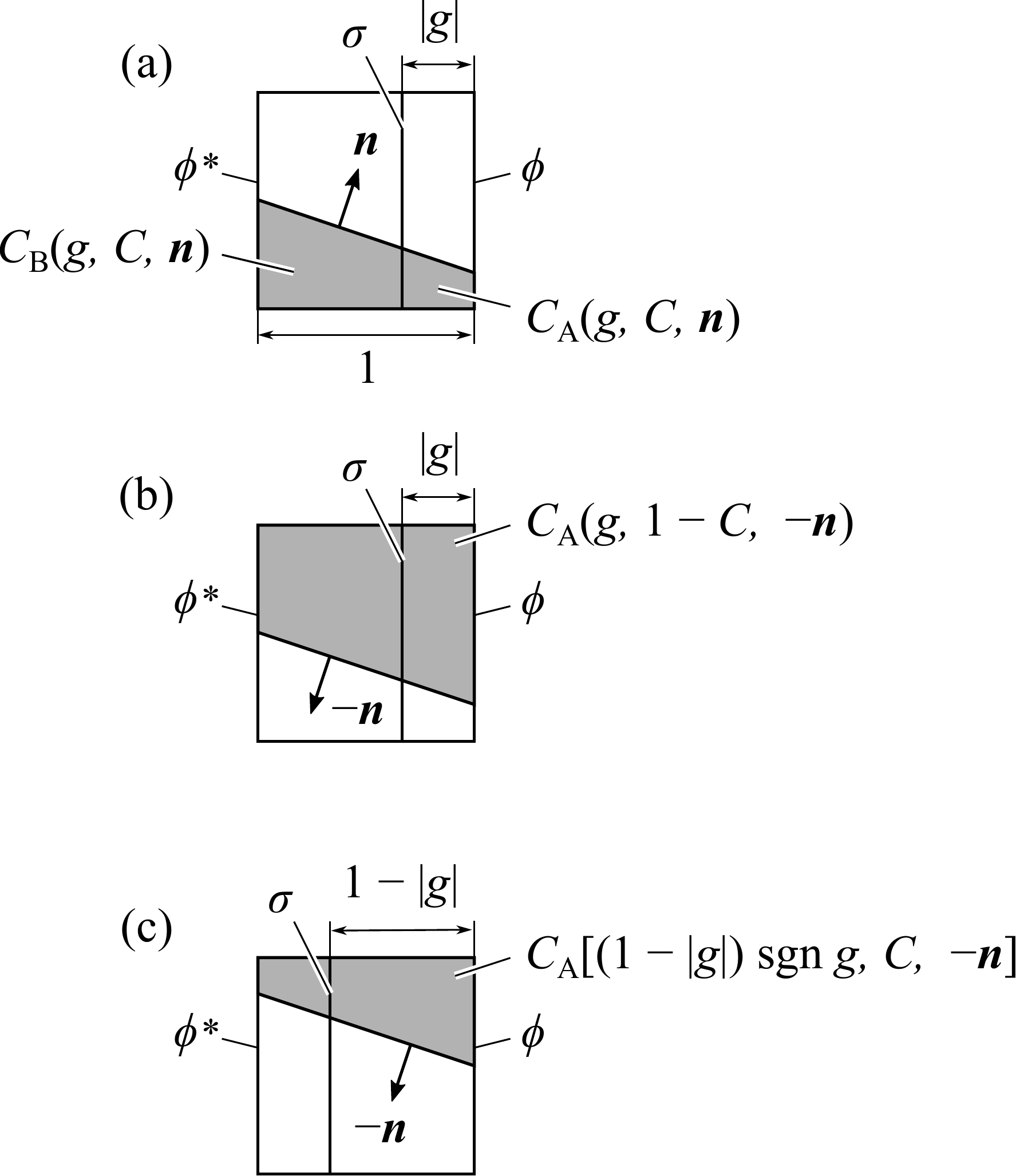}
\caption{Two-dimensional schematic of the relation among partial volume fractions
in a donor cell.
 The sides of the cells are scaled to be unity.}
\label{fig.cacacb}
\end{figure}

The crude APPLIC method is not practical because of a defects described below.
Let us examine whether fluxes evaluated by the crude APPLIC method have
properties which are essential to be satisfied by fluxes of volume fractions.
The computational advection flux $F$
should satisfy the followings conditions:
\begin{align}
F(g, C, \Bold{n}) &
\begin{cases}
\ge 0 & \text{if $g > 0$,}\\
= 0   & \text{if $g = 0$,}\\
\le 0 & \text{if $g < 0$,}
\end{cases}
\label{eq:sgnprop}
\\
F[g, C, (n_1, n_2, n_3)]
&= s_I F[s_I g, C, (s_1n_1, s_2n_2, s_3n_3)],
\label{eq:symm_sign_pn}
\\
F[g, C, (n_1, n_2, n_3)] &= F[g, C, (n_1, n_3, n_2)],
\label{eq:perm_symm_2,3}
\\
g &= F(g, C, \Bold{n}) + F(g, 1-C, -\Bold{n}),
\label{eq:g=FF}
\\
\abs{F(\g, C, \Bold{n})} &\le C,
\label{eq:f_range_loose}
\end{align}
where $s_i$ in Eq.~\eqref{eq:symm_sign_pn} is either $1$ or $-1$.
In Eq.~\eqref{eq:perm_symm_2,3} we suppose that $I=1$ for simplicity.
Condition \eqref{eq:sgnprop} specifies the sign of $F$.
Condition \eqref{eq:symm_sign_pn}
\revised{stems} from
the symmetry of positive and negative directions along the coordinate axes.
Conditions \eqref{eq:perm_symm_2,3}
\revised{stems} from
the permutation symmetry between the second- and the third-coordinate axes.
As shown in (a) and (b) of Fig.~\ref{fig.cacacb},
the second term in the right-hand side of Eq.~\eqref{eq:g=FF}
corresponds to the flux of the light fluid.
Therefore, condition \eqref{eq:g=FF} means that
the total flux, given by $g$, is the sum of the light fluid flux and the dark fluid flux.
Condition \eqref{eq:f_range_loose} provides the upper limit of $\abs{F}$
under the CFL condition.
\revisedBegin

A lower limit of $\abs{F}$ is derived
from Eqs.~\eqref{eq:sgnprop},
\eqref{eq:g=FF} and \eqref{eq:f_range_loose}
as follows.
From Eqs.~\eqref{eq:sgnprop} and \eqref{eq:g=FF},
we have
\begin{align}
\abs{F(g,C,\Bold{n})} +
\abs{F(g,1-C,\Bold{n})} &=\absg,\notag\\
\abs{F(g,1-C,\Bold{n})} &\le 1-C.
\end{align}
These leads
\begin{align}
\abs{F(g,C,n)}\ge \absg - (1-C).\label{neq:lower}
\end{align}
\revisedEnd

\label{def:setS}
Although conditions \eqref{eq:sgnprop}, \eqref{eq:symm_sign_pn},
\eqref{eq:perm_symm_2,3}, and \eqref{eq:g=FF}
are always satisfied by the crude APPLIC method,
conditions \eqref{eq:f_range_loose} \revised{and \eqref{neq:lower} are} not.
To demonstrate this,
a set of sample points, ${S} = \{(\g^{(l)},C^{(l)},\Bold{n}^{(l)}) \,|\, l=1,2,\ldots,N\}$
$(N=10,000,000)$, is used.
The sample points in the set $S$ were
generated such that $\{\g^{(l)}\}$, $\{C^{(l)}\}$, and $\{\Bold{n}^{(l)}\}$
are uniformly distributed on $[0,1]$, $[0, 1]$, and the unit spherical surface, respectively,
by use of pseudorandom numbers.
%
%
Among the sample points in the set $S$, \revised{5.4}\% of the points
do not satisfy the condition \revised{$\absg - (1 - C) \le \abs{\Til{F}_{\rm CAPPLIC}} \le C$},
where $\Til{F}_{\rm CAPPLIC}$ is a flux obtained by the crude APPLIC method.
An easy remedy for the defect is to adopt the limiter as follows:
\revisedBegin
\begin{align}
\Til{F}_{\rm CAPPLIC/L} = \min\{C, \max[\abs{\Til{F}_{\rm CAPPLIC}}, \absg -(1 - C)]\}\sgn g,\label{eq:limiter}
\end{align}
where $\Til{F}_{\rm CAPPLIC/L}$ is a flux obtained by the crude APPLIC method with
the limiter.
\revisedEnd

\subsection{The APPLIC method}

Consider the following relation:
\begin{align}
\Til{C}_{\rA} + \Til{C}_{\rB}&=C\label{eq:defect1},
\end{align}
where $\Til{C}_{\rB}$ is \revised{obtained by}
\begin{align}
\Til{C}_{\rB} (g, C, \Bold{n}) &= \Til{C}_{\rA}[(1 - \absg)\sgn g, C, -\Bold{n}].\label{eq:tcrbgc}
\end{align}
See (a) and (c) of Fig.~\ref{fig.cacacb} for a geometric interpretation of Eq.~\eqref{eq:tcrbgc}.
Equation~\eqref{eq:defect1} is identical to Eq.~\eqref{eq:c_eq_cacb} except that
${C}_{\rA}$ and ${C}_{\rB}$
are obtained by use of the approximation functions $\Til{V}$ and $\Til{\alpha}$.
Generally, Eq.~\eqref{eq:defect1} does not hold because of approximation errors in
$\Til{V}$ and $\Til{\alpha}$.

To improve the crude APPLIC method,
we take advantage of the defect that Eq.~\eqref{eq:defect1} does not hold.
There are two ways to calculate $F$ by use of $\Til{V}$ and $\Til{\alpha}$:
\begin{align}
\Til{F}_\rA(g, C, \Bold{n})&= \Til{C}_\rA(g, C, \Bold{n}) \sgn g,\label{eq:FA}\\
\Til{F}_\rB(g, C, \Bold{n})&= [C-\Til{C}_\rB(g, C, \Bold{n})] \sgn g.
\end{align}
In general, $\Til{F}_\rA$ and $\Til{F}_\rB$  are close but not equal.
The crude APPLIC method uses only Eq.~\eqref{eq:FA} to evaluate flux
(i.e., $\Til{F}_{\rm CAPPLIC}$ is identical to $\Til{F}_\rA$),
whereas the APPLIC method uses either $\Til{F}_\rA$ or $\Til{F}_\rB$
as follows:
\begin{align}
\Til{F}_{\rm APPLIC}(g, C, \Bold{n}) &= \begin{cases}
\Til{F}_\rA(g, C, \Bold{n})&\text{if $\pred(g, C, \Bold{n})$ is true},\\
\Til{F}_\rB(g, C, \Bold{n})&\text{otherwise}.
\end{cases}
\end{align}
where $\pred$ is a choice criterion, which is a logical (or Boolean-valued)
function that returns either a true or false value.
The ideal (i.e., impractical) criterion returns true
if $\abs{\Til{F}_\rA(g, C, \Bold{n}) - F_{\rm PLIC}(g, C, \Bold{n})}$
is smaller than $\abs{\Til{F}_\rB(g, C, \Bold{n}) - F_{\rm PLIC}(g, C, \Bold{n})}$
and false otherwise,
where $F_{\rm PLIC}$ denotes the flux obtained by the PLIC method.

Using Eq.~\eqref{eq_A_set}, we have
\begin{subequations}
\begin{align}
\Til{C}_\rA(g, C, \Bold{n}) &= \absg\, \Til{V}(\Til{\alpha}''_\rA, \Bold{m}''_\rA),\\
\Til{C}_\rB(g, C, \Bold{n}) &= (1-\absg)\, \Til{V}(\Til{\alpha}''_\rB, \Bold{m}''_\rB),
\end{align}
with
\begin{align}
\Til{\alpha}''_\rA&=
\begin{cases}
Q'_\rA \Til{\alpha}'&           \text{if $n_I\, \g  \le 0$},\\
Q'_\rA (\Til{\alpha}' - r_\rA)& \text{if $n_I\, \g > 0$},
\end{cases}
\\
\Til{\alpha}''_\rB&=
\begin{cases}
Q'_\rB (\Til{\alpha}' - r_\rB)& \text{if $n_I\, \g \le 0$},\\
Q'_\rB \Til{\alpha}'&           \text{if $n_I\, \g  > 0$},
\end{cases}
\\
m''_{\rB,l} &=
\begin{cases}
Q'_\rB m'_l (1-\absg) & \text{if $l=I$},\\
Q'_\rB m'_l & \text{if $l\ne I$},
\end{cases}
\\
Q'_\rB &= \frac{1}{1 - r_\rB},
\\
r_\rB &= m'_I\,\absg,
\\
\Til{\alpha}'&=\Til{\alpha}(C, \Bold{m}).
\end{align}
\end{subequations}

The author proposes the following choice criterion for the APPLIC method:
true if $\abs{\Til{\alpha}''_\rA-1/2} > \abs{\Til{\alpha}''_\rB-1/2}$
and false otherwise.
See Appendix B for the derivation of the choice criterion.
Figure \ref{fig.applic} shows an example implementation of the APPLIC method with the
proposed criterion written in Fortran 90.
\begin{figure}[H]
\begin{lstlisting}
  function calc_flux_applic(g, c, vn1, vn2, vn3) result(f)
    ! Preconditions:$\;{\tt g}\in(-1,1),\;{\tt c}\in(0,1).$
    use constants
    real(8), intent(in) :: g, c, vn1, vn2, vn3
    real(8) :: f, absg, alpha, alphb, ra, rb, qa, qb, vm1, vm2, vm3, v, sw
    absg = abs(g)
    vm1 = abs(vn1)
    vm2 = abs(vn2)
    vm3 = abs(vn3) + CONST_TINY
    qa = 1d0 / (vm1 + vm2 + vm3)
    vm1 = vm1 * qa
    vm2 = vm2 * qa
    vm3 = vm3 * qa
    alpha = calc_approx_alpha(c, vm1, vm2, vm3)
    rb = vm1 * absg
    ra = vm1 - rb
    qb = 1d0 / (1d0 - rb)
    qa = 1d0 / (1d0 - ra)
    alphb = (alpha - merge(0d0, rb, vn1 * g > 0d0)) * qb
    alpha = (alpha - merge(ra, 0d0, vn1 * g > 0d0)) * qa
    sw = 0d0
    if (abs(alphb - 0.5d0) > abs(alpha - 0.5d0)) then
      sw = 1d0
      alpha = alphb
      qa = qb
      rb = ra
    end if
    v = calc_approx_v(alpha, rb * qa, vm2 * qa, vm3 * qa)
    f = sign((c - v) * sw + v * absg, g)
  end function calc_flux_applic
\end{lstlisting}
\caption{Example implementation of the APPLIC method with the
proposed criterion written in Fortran 90.}
\label{fig.applic}
\end{figure}

Now the approximation accuracy of fluxes obtained by the APPLIC method is examined.
Table~\ref{tbl_correct_answer_rates} compares the statistics for the approximation errors
of the crude APPLIC method,
crude APPLIC method with Eq.~\eqref{eq:limiter} as a limiter,
APPLIC method, and APPLIC method with the ideal criterion,
for the point set $S$.
Here, the approximation errors of a flux
is defined as the difference between the flux and that obtained by
the PLIC method.
It is observed that applying the limiter to the crude APPLIC method
reduces the mean error \revised{and the maximum error}.
\revised{The mean error of the APPLIC method is smaller than the crude APPLIC method with the limiter.}
The last row is for the APPLIC method with the ideal
criterion, and indicates
the lower limits for the mean errors and the maximum errors
for any choice criteria.
\begin{table}[H]
\centering
\caption{
Statistics of the approximation errors of fluxes for the point set $S$.
}
\begin{tabular}{cccc}
\hline
Method & Correct answer rate$^{\rm a}$ & Mean error$^{\rm b}$ & Maximum error$^{\rm c}$\\
\hline
Crude APPLIC                       & 50.0\% & $2.89\times 10^{-3}$ & $\revised{3.32}\times 10^{-2}$\\
Crude APPLIC with limiter      & ---    & $\revised{2.54}\times 10^{-3}$ & $\revised{2.77}\times 10^{-2}$\\
APPLIC                             & 71.5\% & $1.88\times 10^{-3}$ & $\revised{2.77}\times 10^{-2}$\\
APPLIC with ideal criterion    & 100\%  & $1.43\times 10^{-3}$ & $\revised{1.67}\times 10^{-2}$\\
\hline
\multicolumn{4}{p{37em}}%
{\footnotesize $\!\!\!^{\rm a}$The percentage of the sample points where the criterion
provides true if $\abs{\Til{F}_{\rA}-F_{\rm PLIC}} < \abs{\Til{F}_{\rB}-F_{\rm PLIC}}$
and false otherwise.}\\
\multicolumn{4}{p{37em}}%
{\footnotesize $\!\!\!^{\rm b}$The arithmetic mean of the absolute errors.}\\
\multicolumn{4}{p{37em}}%
{\footnotesize $\!\!\!^{\rm c}$The maximum value of the absolute errors.}
\end{tabular}
\label{tbl_correct_answer_rates}
\end{table}

\begin{figure}[H]
\centering
\includegraphics[width=8.5cm]{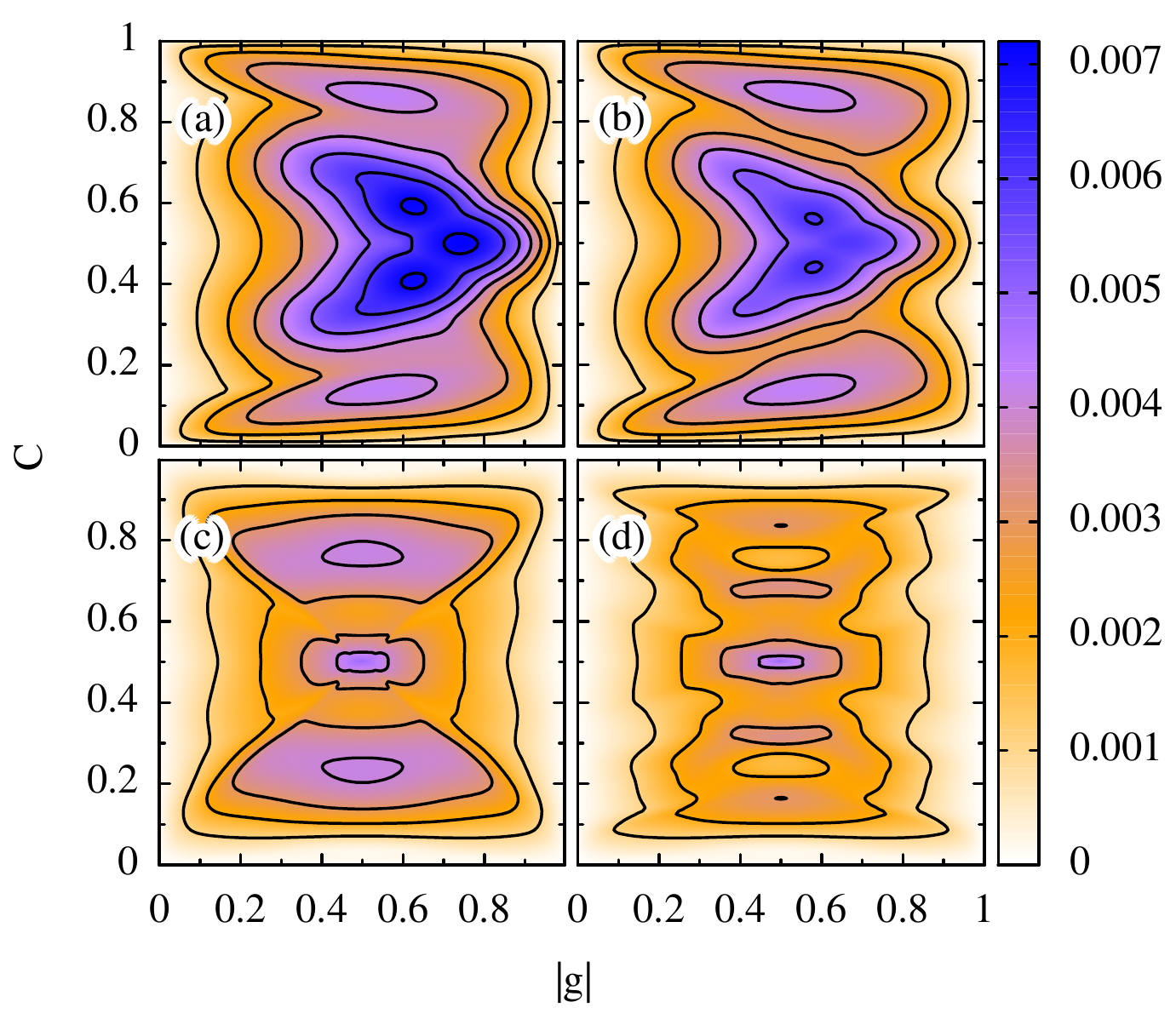}
\caption{\revised{Dependence of the arithmetic means of absolute approximation errors
in fluxes $\Til{F}(g, C, \Bold{n})$ on $\absg$ and $C$, obtained by
(a) crude APPLIC, (b) crude APPLIC with limiter,
(c) APPLIC, and (d) APPLIC with ideal criterion.
The contour lines are drawn at 0.001, 0.002, 0.003, $\ldots$, 0.007.}}
\label{fig.meanerror}
\end{figure}

\begin{figure}[H]
\centering
\includegraphics[width=8.5cm]{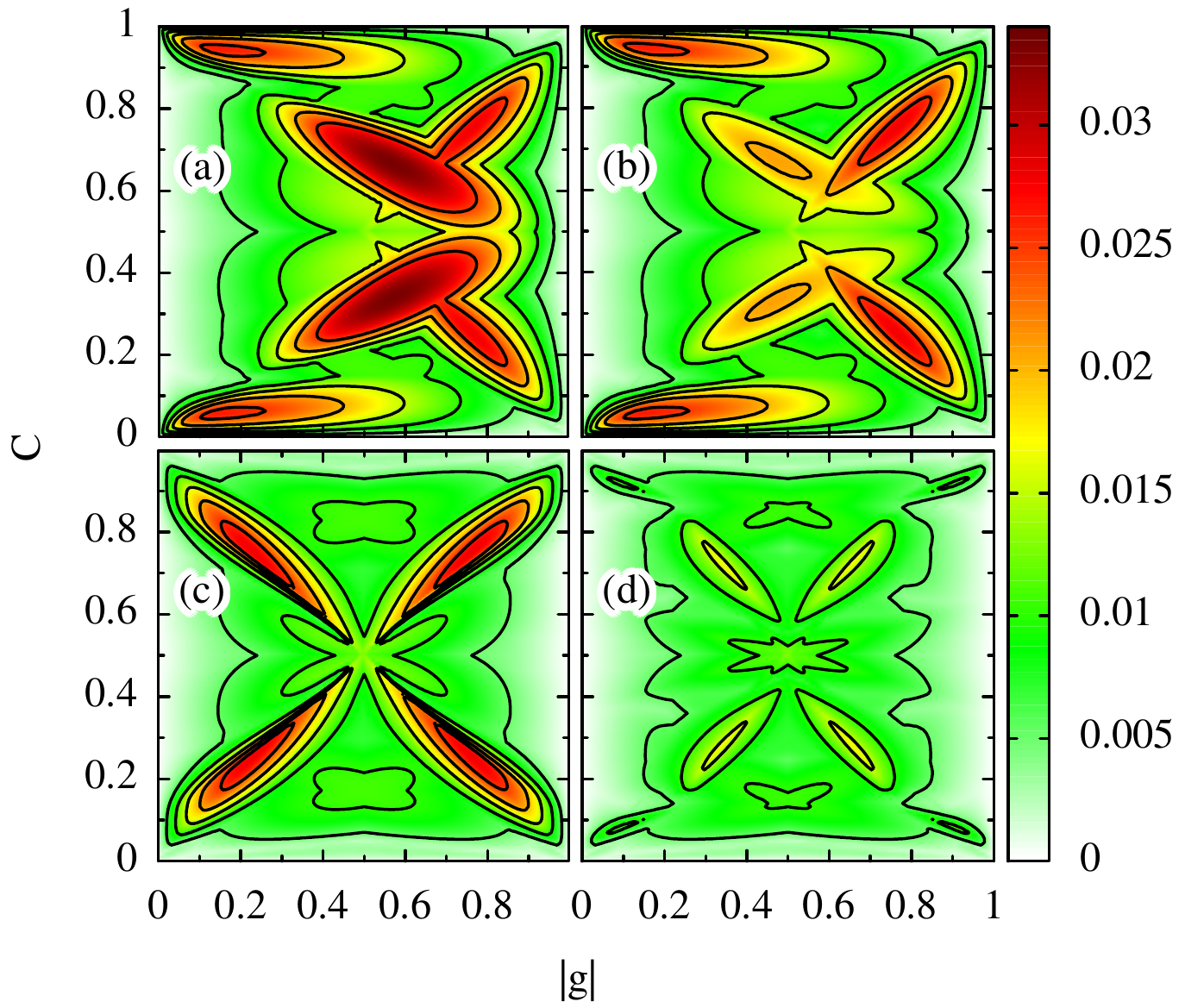}
\caption{\revised{Dependence of the maximum values of absolute approximation errors in fluxes
$\Til{F}(g, C, \Bold{n})$ on $\absg$ and $C$, obtained by
(a) crude APPLIC, (b) crude APPLIC with limiter,
(c) APPLIC, and (d) APPLIC with ideal criterion.
The contour lines are drawn at 0.005, 0.01, 0.015, $\ldots$, 0.03.}}
\label{fig.maxerror}
\end{figure}

\revisedBegin
Figures \ref{fig.meanerror} and \ref{fig.maxerror}
show the dependence of approximation errors of fluxes on $\absg$ and $C$.
The errors in the figures were evaluated by using
50000 three-dimensional vectors
distributed uniformly on the unit spherical surface
as sample points of $\Bold{n}$.
All the plots are axisymmetric with respect to $C = 0.5$
since the PLIC, crude APPLIC, crude APPLIC with limiter,
APPLIC, and APPLIC with ideal criterion satisfy Eq.~\eqref{eq:g=FF}.
Moreover, the plots for the APPLIC and the APPLIC with ideal criterion
are axisymmetric with respect to $\absg = 0.5$
since the PLIC, APPLIC,
and APPLIC with ideal criterion satisfy that
$F_{\rA}$ equals to $F_{\rB}$. 
\revisedEnd


The APPLIC method satisfies conditions \eqref{eq:f_range_loose} \revised{and \eqref{neq:lower}} as well as
conditions \eqref{eq:sgnprop}-\eqref{eq:g=FF}.
Moreover, the following condition tighter than conditions \eqref{eq:sgnprop},
 \eqref{eq:f_range_loose}\revised{, and \eqref{neq:lower}}
is also satisfied:
\begin{align}
F\in[{\rm LB}, {\rm UB}],\label{eq:lbub}
\end{align}
with
\begin{align}
{\rm LB}&= \begin{cases}
g C                   &\text{if $n_I \le 0$,}\\
\max[g - (1 - C), 0]  &\text{if $n_I > 0$ and $g \ge 0$,}\\
\max[g, -C]           &\text{if $n_I > 0$ and $g <   0$,}
\end{cases}
\\
{\rm UB}&=
\begin{cases}
g C                   &\text{if $n_I \ge 0$,}\\
\min[g + (1 - C), 0]  &\text{if $n_I < 0$ and $g \le 0$,}\\
\min[g, C]            &\text{if $n_I < 0$ and $g >   0$.}
\end{cases}
\end{align}
These bounds are derived from fluxes for SLIC-type fluid configurations~\cite{marek2008}.

We now demonstrate that the APPLIC method satisfies condition \eqref{eq:lbub}
using the set $S$.
Among the sample points in the set $S$, 5.6\% of the points do not satisfy
the relation $\Til{F}_\rA \in [{\rm LB}, {\rm UB}]$.
Similarly, 5.6\% of the points do not satisfy $\Til{F}_\rB \in[{\rm LB}, {\rm UB}]$.
However, all the points meet either
$\Til{F}_\rA\in[{\rm LB}, {\rm UB}]$ or $\Til{F}_\rB\in[{\rm LB}, {\rm UB}]$.

\begin{figure}[H]
\centering
\includegraphics[width=9cm]{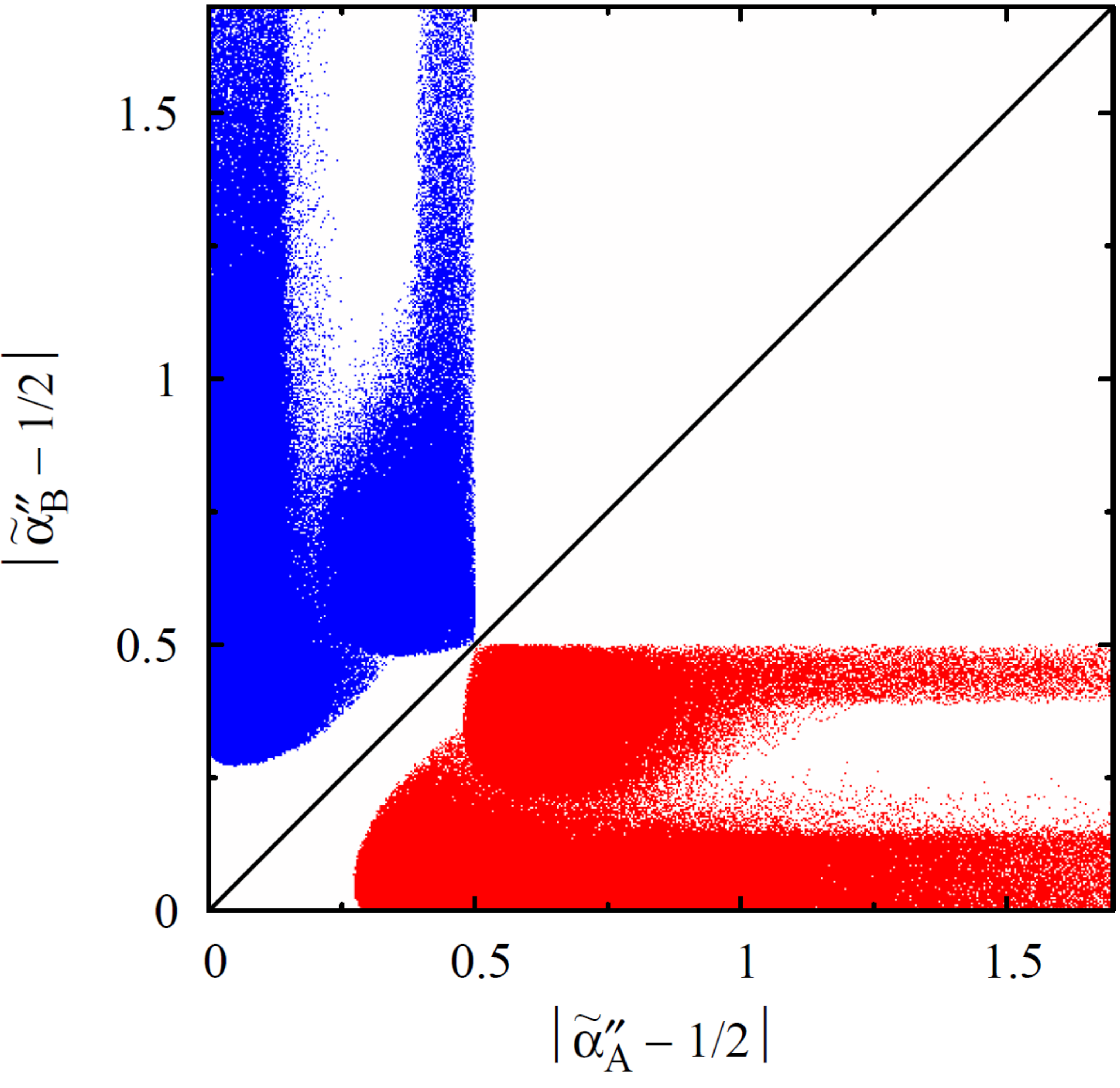}
\caption{The sample points in the set $S$ so that
${\Til{F}_\rA} \notin[\rm LB,UB]$ (blue dots) and
${\Til{F}_\rB} \notin[\rm LB,UB]$ (red dots).
The solid line represents the line $\abs{\Til{\alpha}''_\rA-1/2}=\abs{\Til{\alpha}''_\rB-1/2}$.}
\label{fig_improper_points}
\end{figure}

Figure~\ref{fig_improper_points} depicts the points in the set $S$ such that
${\Til{F}_\rA} \notin[{\rm LB}, {\rm UB}]$ (blue dots) and
${\Til{F}_\rB} \notin[{\rm LB}, {\rm UB}]$ (red dots).
Each dot is placed at $(\abs{\Til{\alpha}''_\rA-1/2},\,\abs{\Til{\alpha}''_\rB-1/2})$ on the plot.
All the blue and the red dots lie in the regions
$\abs{\Til{\alpha}''_\rA-1/2} < \abs{\Til{\alpha}''_\rB-1/2}$ and
$\abs{\Til{\alpha}''_\rA-1/2} > \abs{\Til{\alpha}''_\rB-1/2}$, respectively.
This indicates that fluxes evaluated by
the APPLIC method with the proposed choice criterion always satisfy Eq.~\eqref{eq:lbub}.

\section{Numerical tests}
\subsection{The accuracy of advection}
We compare the accuracy of advection for the APPLIC method
with other VOF methods over three test problems.
The VOF methods examined in this section are as follows:
SVOF,  VOF/WLIC, THINC/SW,  THINC/WLIC,
 PLIC, and  APPLIC.
Each test problem is designed so that
the initial distribution (at $t=0$) of the light and dark fluids is
theoretically identical with the final distribution.
The $L_1$ error, defined as
${\sum_{i,j,k}(\Delta x)^3\left\lvert C_{i,j,k}(t=0) - C_{i,j,k}(t=T) \right\rvert}$,
is employed to compare the accuracy,
where $T$ is the final time.
Parker and Youngs' method \cite{scardovelli2003interface,parker1992two} is adopted
to evaluate surface normals.
We adopt
an operator splitting algorithm for the advection of
volume fractions as follows:
\begin{subequations}
\begin{align}
C_{i,j,k}^*&=C_{i,j,k}^{(n)}
-F_{i+\Half,j,k}^{(n)}+F_{i-\Half,j,k}^{(n)}
+\frac{C_{i,j,k}^{(n)}\Delta t}{\Delta x}
(u_{1:i+\Half,j,k} - u_{1:i-\Half,j,k}),\\
C_{i,j,k}^{**}&=C_{i,j,k}^*
-F_{i,j+\Half,k}^*+F_{i,j-\Half,k}^*
+\frac{C_{i,j,k}^{(n)}\Delta t}{\Delta x}
(u_{2:i,j+\Half,k} - u_{2:i,j-\Half,k}),\\
C_{i,j,k}^{(n+1)}&=C_{i,j,k}^{**}
-F_{i,j,k+\Half}^{**}+F_{i,j,k-\Half}^{**}
+\frac{C_{i,j,k}^{(n)}\Delta t}{\Delta x}
(u_{3:i,j,k+\Half} - u_{3:i,j,k-\Half}),
\end{align}
\label{eq:sweep}
\end{subequations}
where the superscript $(n)$ refers to the temporal indices,
and the superscripts $*$ and $**$ represent quantities
at the first and second intermediate steps,
respectively.
The order of directions of Eq.~\eqref{eq:sweep}
is changed at each time step to minimize
possible asymmetries.
All floating point arithmetic is done in double-precision.

\subsubsection{\revised{Test 1}}
In the first test problem,
a shape
defined as the union
of the rectangular parallelepipeds
$\{\Bold{x}\in[0.08, 0.48]\times[0.2, 0.36]\times[0.2, 0.36]\}$
and the sphere with center $(0.28, 0.28, 0.28)$ and radius $0.15$
is translated
in a computational domain $[0,1]\times[0,1]\times[0,1]$.
We set the final time as $T = 0.8$.
The shape $B$ is advected in the following uniform velocity field:
\begin{align}
\Bold{u}=\begin{cases}
(1, 1, 1)& \text{if $t < T/2$},\\
(-1, -1, -1)& \text{if $t > T/2$}.
\end{cases}
\end{align}

\begin{figure}[H]
\centering
\includegraphics[width=12cm]{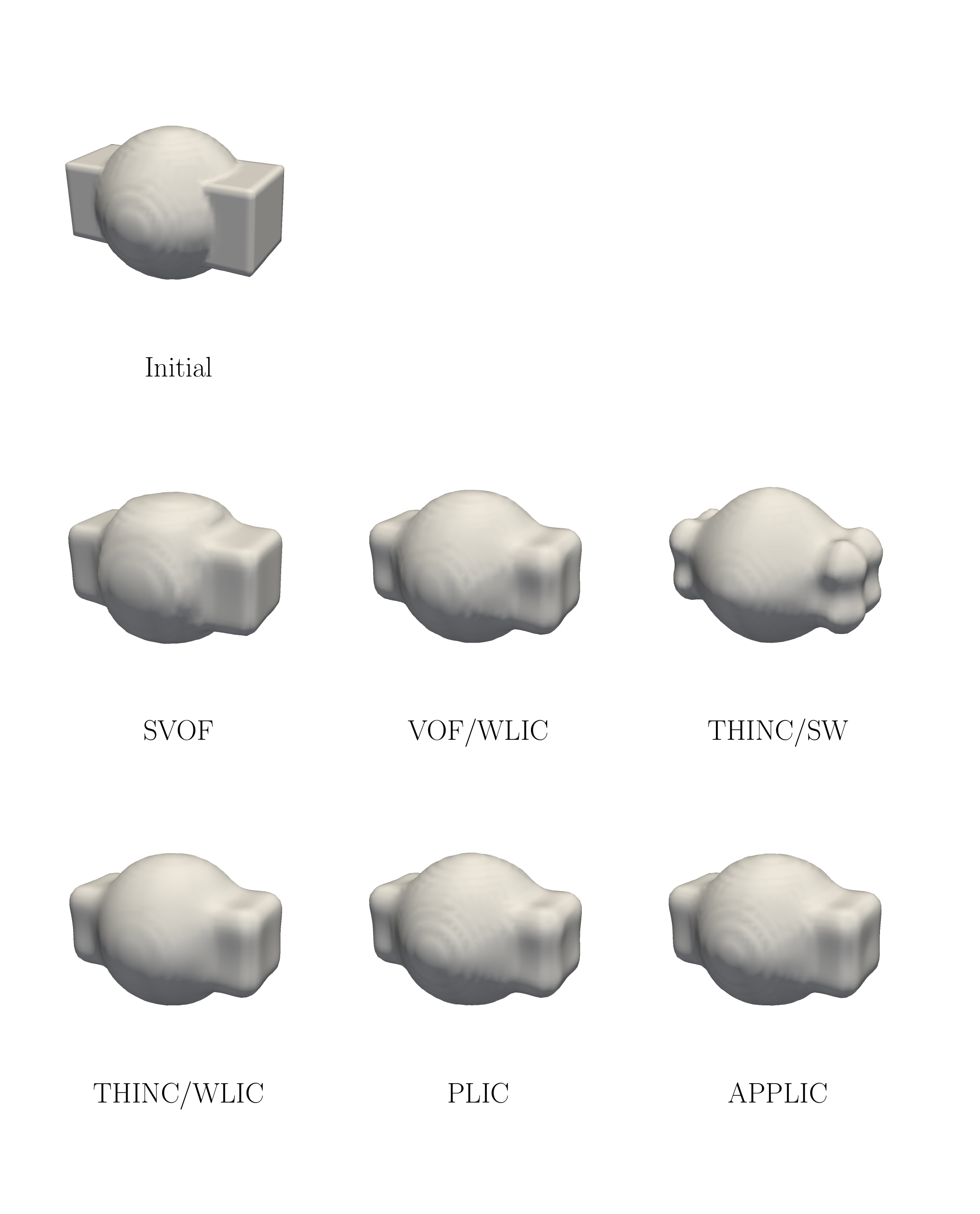}
\caption{The initial $(t=0)$ and the final $(t=T)$ shapes
for test 1,
using a 100$\times$100$\times$100 grid with $\Delta t=0.005$.}
\label{fig_test1}
\end{figure}
\begin{figure}[H]
\centering
\includegraphics[width=12cm]{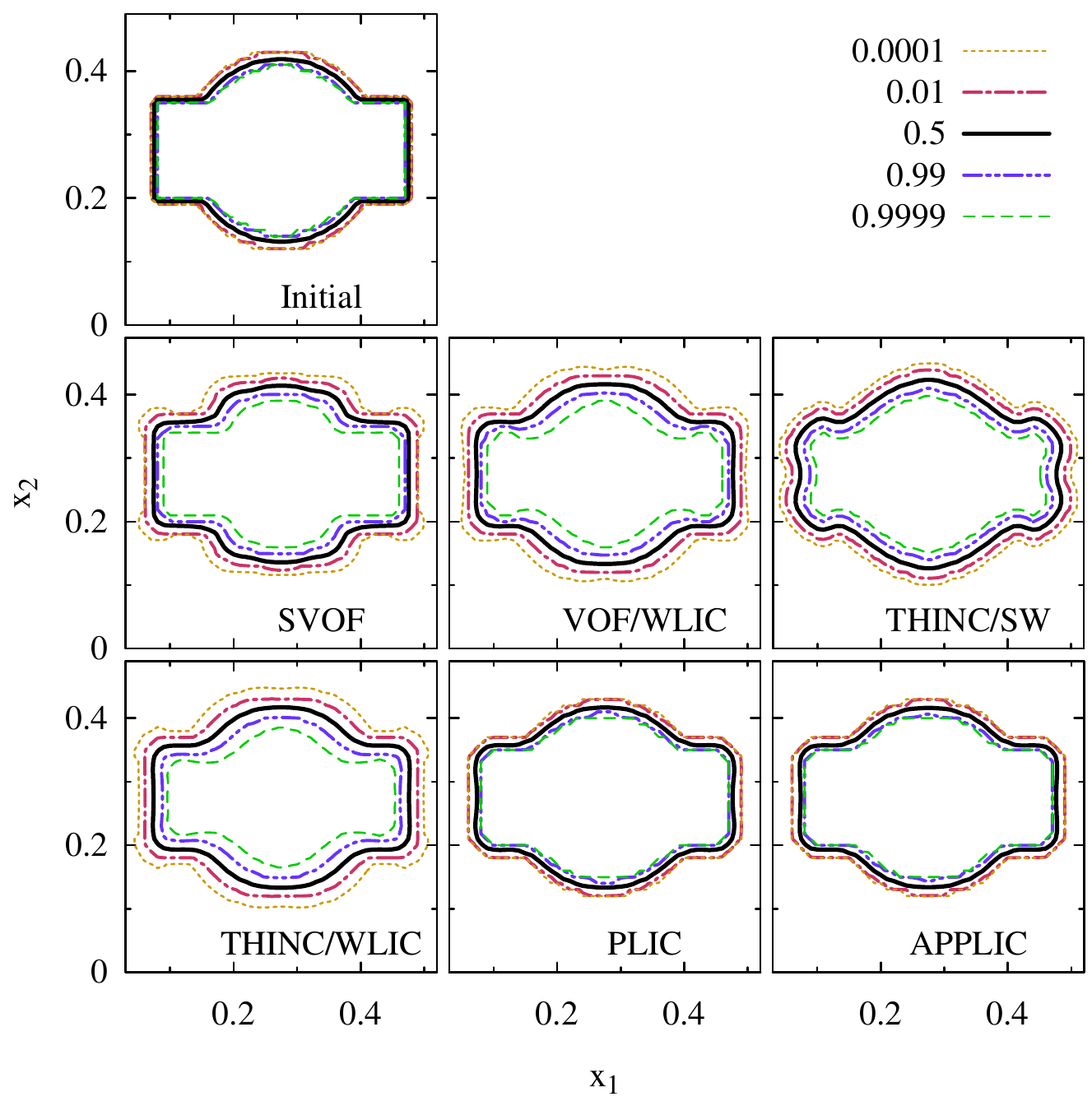}
\caption{\revised{Slice plots of the initial $(t=0)$ and final $(t=T)$ volume fractions
for test 1 at $x_3=0.24$,
using a $100\times100\times100$ grid with $\Delta t=0.005$.
Volume fractions are drawn by contour lines of 0.00001, 0.01, 0.5, 0.99, and 0.99999.}}
\label{fig_test1_cntr}
\end{figure}

Figure~\ref{fig_test1} compares the initial shape and
the final shapes
using a $100\times 100 \times 100$ ($\Delta x = 0.01$) grid with $\Delta t = 0.005$,
where the CFL number is 0.5.
The THINC/SW method produces significantly deformed shape.
\revised{
Sliced plots of volume fractions are shown in Fig.~\ref{fig_test1_cntr}.
This figure shows that
the PLIC and the APPLIC methods keep the interface widths more compact
than the SVOF, VOF/WLIC, THINC/SW, and THINC/WLIC methods.
}
Table \ref{tbl_test1_1} presents the $L_1$ errors and the corresponding convergence rate
for
$ 25\times  25\times  25$ $(\Delta x = 0.04)$,
$ 50\times  50\times  50$ $(\Delta x = 0.02)$, and
$100\times 100\times 100$ $(\Delta x = 0.01)$, grids,
where $\Delta t= 0.02$, 0.001, and 0.005, respectively.
\revised{
The errors of the PLIC and the APPLIC methods are the lowest
and the second lowest for all cases.
}
\begin{table}[H]
\centering
\caption{
$L_1$ errors and convergence rates
using different grid sizes
for test 1.
}
\label{tbl_test1_1}
\begin{tabular}{cccccc}
\hline
Method
 & $ 25 \times 25 \times 25 $  & Rate  & $ 50 \times 50 \times 50 $  & Rate  & $ 100 \times 100 \times 100 $ \\
\hline
SVOF  & $4.19 \times 10^{-3}$& 0.74 & $2.50 \times 10^{-3}$& 1.03 & $1.22 \times 10^{-3}$\\
VOF/WLIC  & $3.77 \times 10^{-3}$& 0.84 & $2.11 \times 10^{-3}$& 1.23 & $8.99 \times 10^{-4}$\\
THINC/SW  & $3.35 \times 10^{-3}$& 0.46 & $2.44 \times 10^{-3}$& 1.03 & $1.19 \times 10^{-3}$\\
THINC/WLIC  & $4.57 \times 10^{-3}$& 0.86 & $2.51 \times 10^{-3}$& 1.17 & $1.12 \times 10^{-3}$\\
PLIC  & $2.71 \times 10^{-3}$& 0.69 & $1.68 \times 10^{-3}$& 1.15 & $7.58 \times 10^{-4}$\\
APPLIC  & $2.81 \times 10^{-3}$& 0.67 & $1.77 \times 10^{-3}$& 1.17 & $7.87 \times 10^{-4}$\\
\hline
\end{tabular}
\end{table}


\subsubsection{\revised{Test 2}}
\begin{figure}[H]
\centering
\includegraphics[width=12cm]{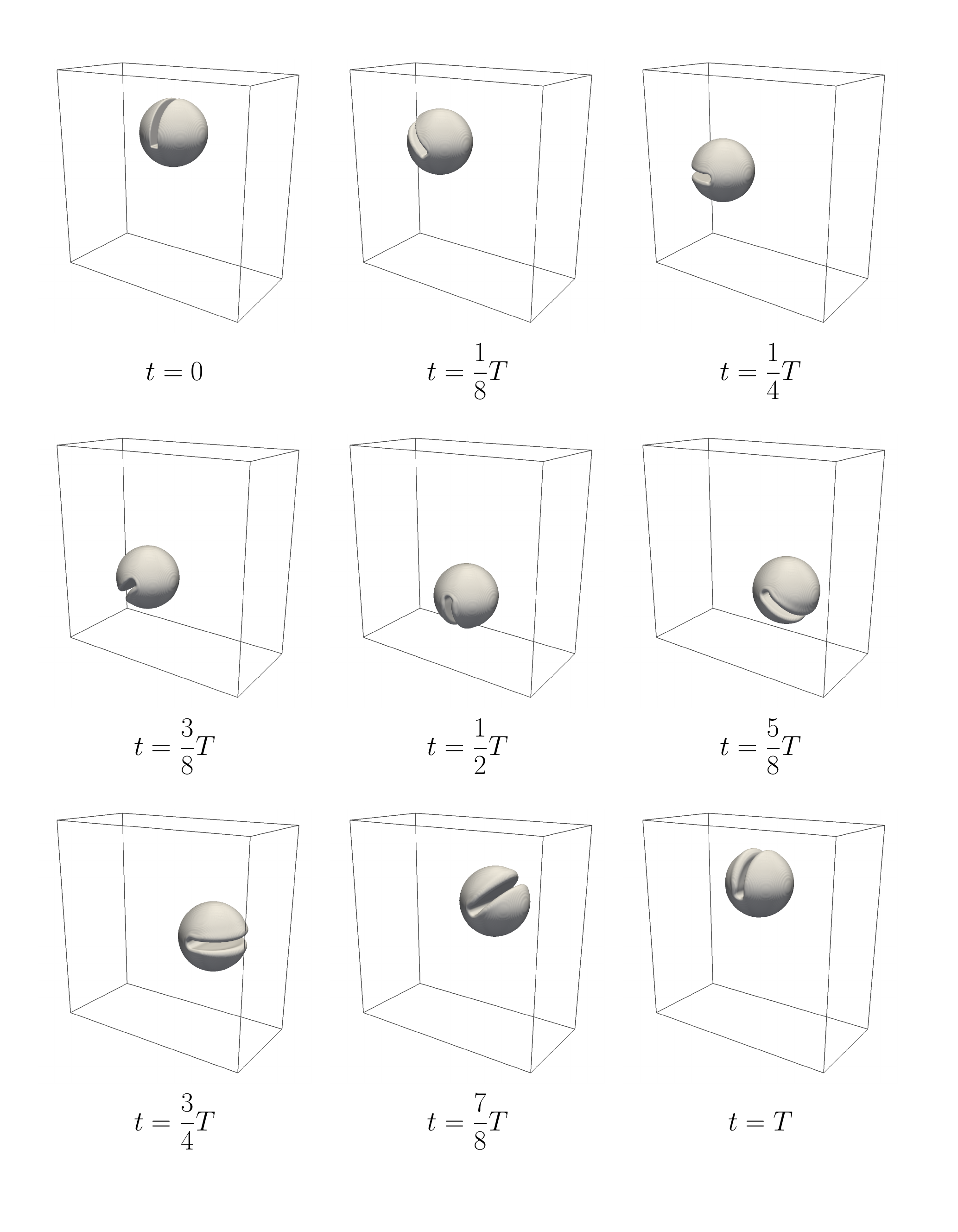}
\caption{Evolution of test 2,
calculated using the APPLIC method on a
$200\times 200\times 96$ grid with $\Delta t = 0.005$.}
\label{fig_test2_evol}
\end{figure}

Secondly, we examine a test problem proposed by Enright et al. \cite{enright2002hybrid},
which is analogous with Zalesak's disk problem in two-dimensional space \cite{zalesak1979fully}.
A sphere of 0.16 radius with a slot of 0.04 wide
and 0.2 deep is located initially at $(0.5, 0.72, 0.24)$
in a computational domain $[0,1]\times[0,1]\times[0,0.48]$
and undergoes a rigid body rotation.
The velocity field is static and is represented by
\begin{align}
u_1 &= (2\pi/T)(0.5-x_2),\\
u_2 &= (2\pi/T)(x_1-0.5),\\
u_3 &= 0.
\end{align}
We set the final time as $T = 6$.
Figure \ref{fig_test2_evol} shows the evolution of the problem.

\begin{figure}[H]
\centering
\includegraphics[width=12cm]{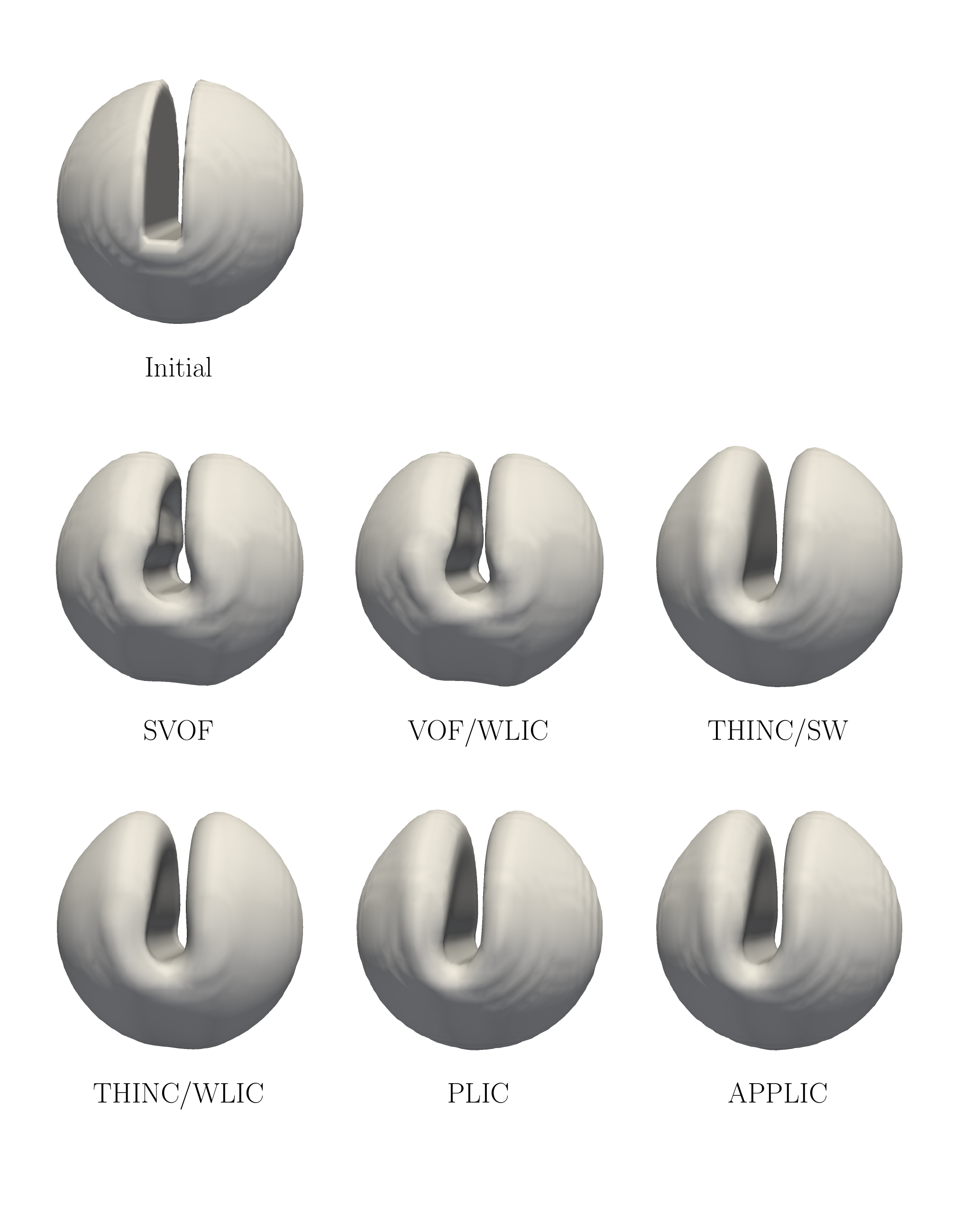}
\caption{The initial $(t=0)$ and the final $(t=T)$ shapes for test 2,
using a 100$\times$100$\times$48 grid with $\Delta t=0.01$.}
\label{fig_test2}
\end{figure}

\begin{figure}[H]
\centering
\includegraphics[width=12cm]{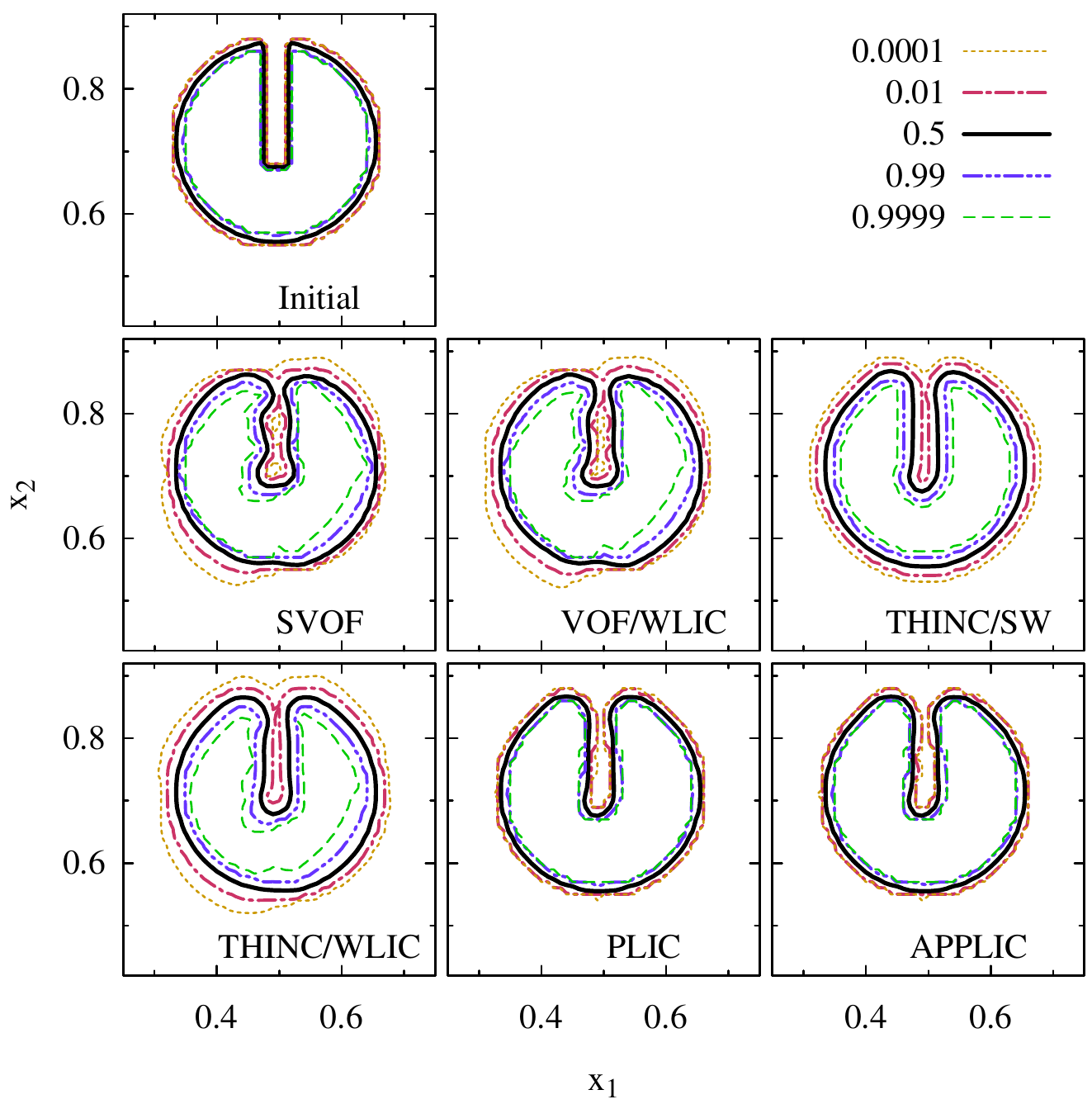}
\caption{\revised{Slice plots of the initial $(t=0)$ and final $(t=T)$ volume fractions
for test 2 at $x_3=0.24$,
using a $100\times100\times48$ grid with $\Delta t=0.01$.
Volume fractions are drawn by contour lines of 0.00001, 0.01, 0.5, 0.99, and 0.99999.}}
\label{fig_test2_cntr}
\end{figure}

Figure \ref{fig_test2} compares the initial (exact) shape and
the final shapes
using a $100\times 100 \times 48$ ($\Delta x = 0.01$) grid and $\Delta t = 0.01$,
where the maximum CFL number is approximately 0.5.
The SVOF and VOF/WLIC methods produces significantly deformed shapes.
\revised{
Sliced plots of volume fractions are shown in Fig.~\ref{fig_test2_cntr}.
This figure shows that
the PLIC and the APPLIC methods keep the interface widths more compact
than the SVOF, VOF/WLIC, THINC/SW, and THINC/WLIC methods.
}
Table \ref{tbl_test2_1}  presents the $L_1$ errors and the corresponding
convergence rates for
$ 25\times  25\times 12$ $(\Delta x = 0.04)$,
$ 50\times  50\times 24$ $(\Delta x = 0.02)$, and
$100\times 100\times 48$ $(\Delta x = 0.01)$, grids,
where $\Delta t= 0.04$, 0.02, and 0.01, respectively.
\revised{The errors of the APPLIC and the PLIC are the smallest
and the second smallest for all cases.}
\begin{table}[H]
\centering
\caption{
$L_1$ errors and convergence rates
using different grid sizes
for test 2.
}
\label{tbl_test2_1}
\begin{tabular}{cccccc}
\hline
Method
 & $ 25 \times 25 \times 12 $  & Rate  & $ 50 \times 50 \times 24 $  & Rate  & $ 100 \times 100 \times 48 $ \\
\hline
SVOF  & $5.25 \times 10^{-3}$& 0.63 & $3.40 \times 10^{-3}$& 1.32 & $1.36 \times 10^{-3}$\\
VOF/WLIC  & $5.39 \times 10^{-3}$& 0.57 & $3.63 \times 10^{-3}$& 1.35 & $1.42 \times 10^{-3}$\\
THINC/SW  & $4.42 \times 10^{-3}$& 0.67 & $2.79 \times 10^{-3}$& 1.50 & $9.84 \times 10^{-4}$\\
THINC/WLIC  & $5.51 \times 10^{-3}$& 0.71 & $3.36 \times 10^{-3}$& 1.30 & $1.37 \times 10^{-3}$\\
PLIC  & $4.00 \times 10^{-3}$& 1.07 & $1.91 \times 10^{-3}$& 1.58 & $6.39 \times 10^{-4}$\\
APPLIC  & $3.99 \times 10^{-3}$& 0.98 & $2.02 \times 10^{-3}$& 1.59 & $6.73 \times 10^{-4}$\\
\hline
\end{tabular}
\end{table}


\subsubsection{\revised{Test 3}}
\begin{figure}[H]
\centering
\includegraphics[width=12cm]{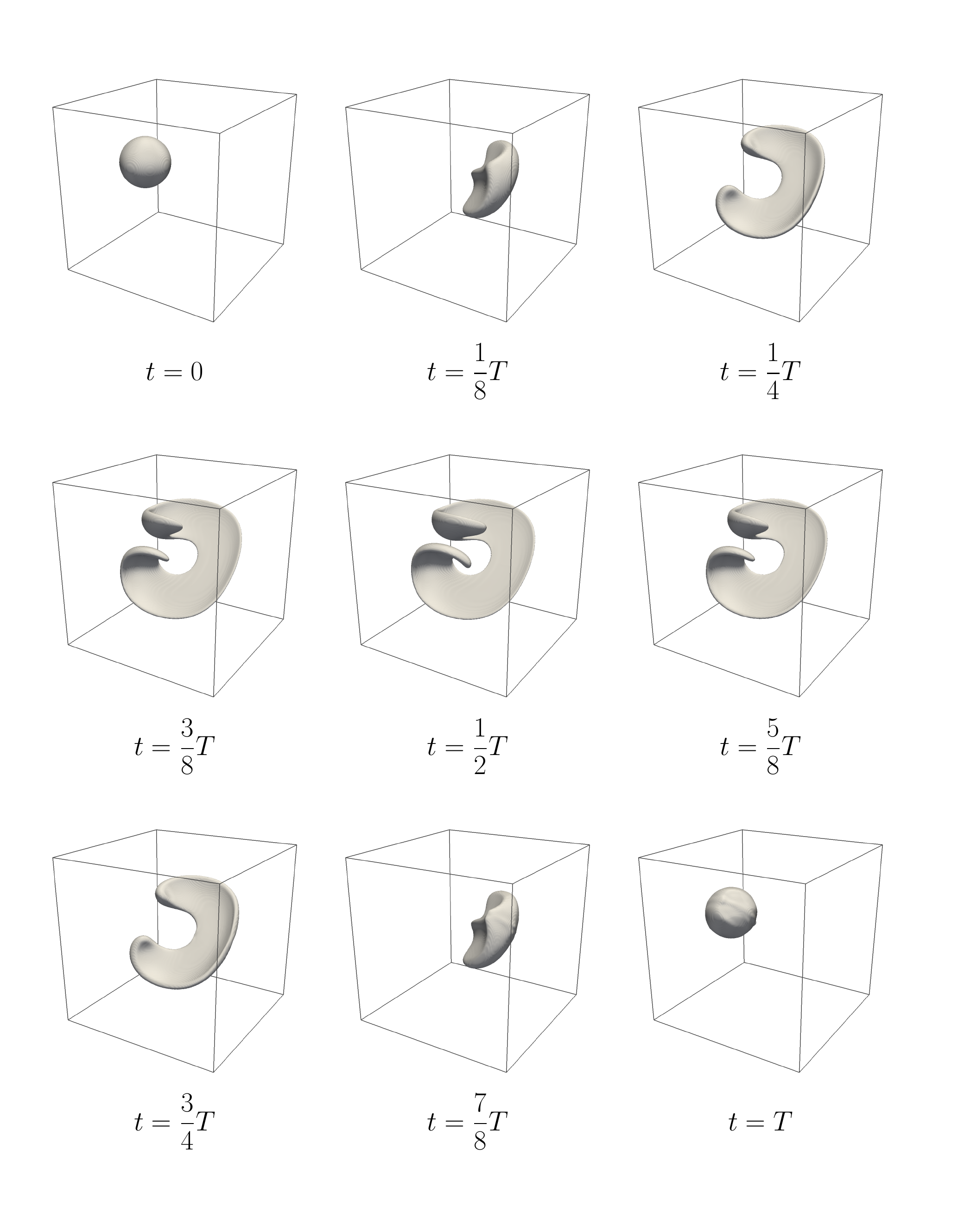}
\caption{Evolution of test 3,
calculated using the APPLIC method on a $200\times 200\times 200$ grid
with $\Delta t = 0.00125$.}
\label{fig_test3_evol}
\end{figure}

In the third test problem,
a sphere of 0.15 radius
centered at $(0.35, 0.35, 0.35)$
in a computational domain $[0,1]\times[0,1]\times[0,1]$
 is deformed in an incompressible flow field proposed by LeVeque \cite{leveque1996high},
expressed by
\begin{align}
u_1 &= 2\sin^2(\pi x_1)\sin (2\pi x_2) \sin (2\pi x_3) \cos(\pi t/T),\\
u_2 &= -\sin (2\pi x_1)\sin^2(\pi x_2) \sin (2\pi x_3) \cos(\pi t/T),\\
u_3 &= -\sin (2\pi x_1)\sin (2\pi x_2) \sin^2(\pi x_3) \cos(\pi t/T).
\end{align}
We set $T=3$.
Figure \ref{fig_test3_evol} shows the evolution of the problem.

\begin{figure}[H]
\centering
\includegraphics[width=12cm]{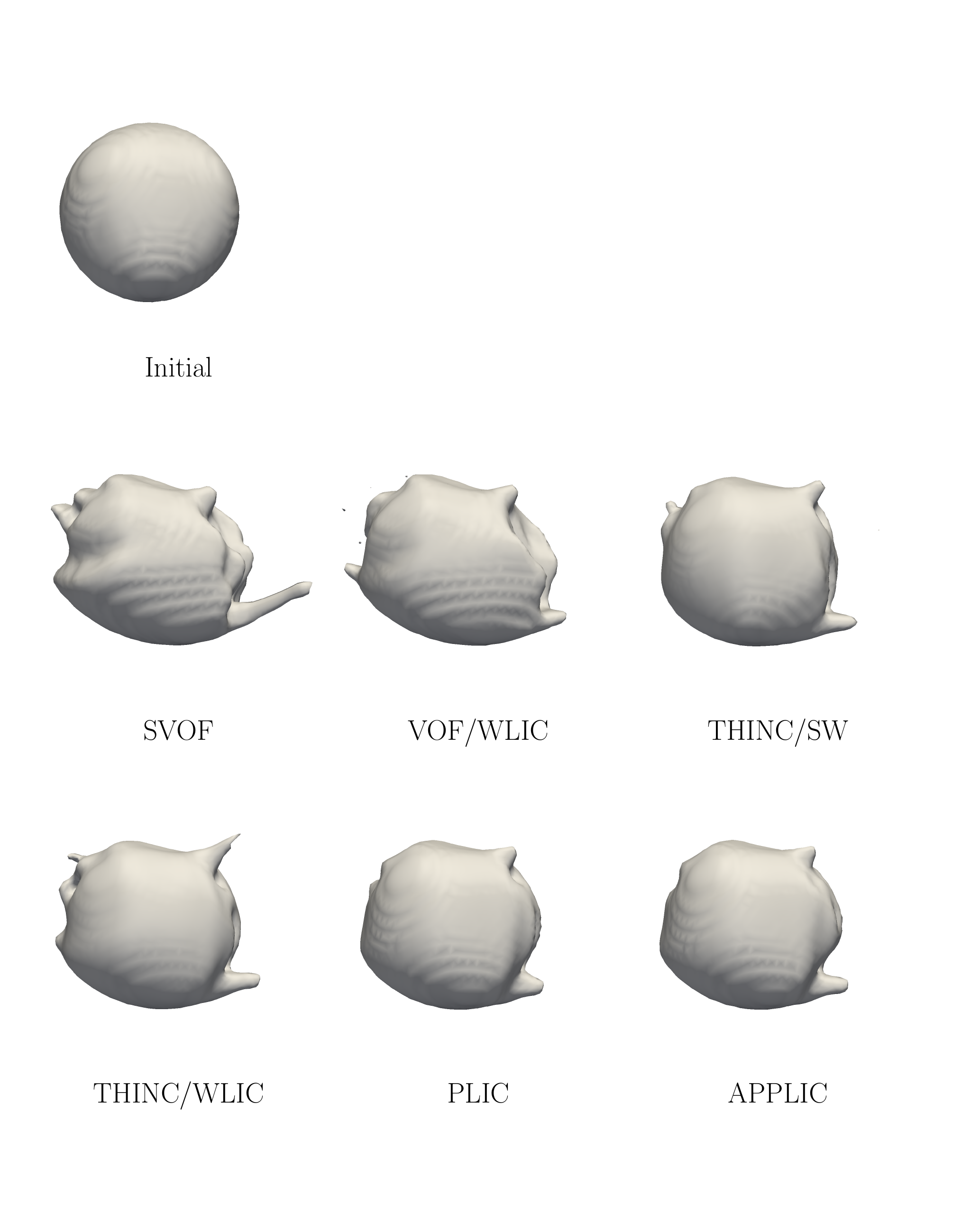}
\caption{The initial $(t=0)$ and the final $(t=T)$ shapes
for test 3,
using a 100$\times$100$\times$100 grid with $\Delta t=0.0025$.}
\label{fig_test3}
\end{figure}

\begin{figure}[H]
\centering
\includegraphics[width=12cm]{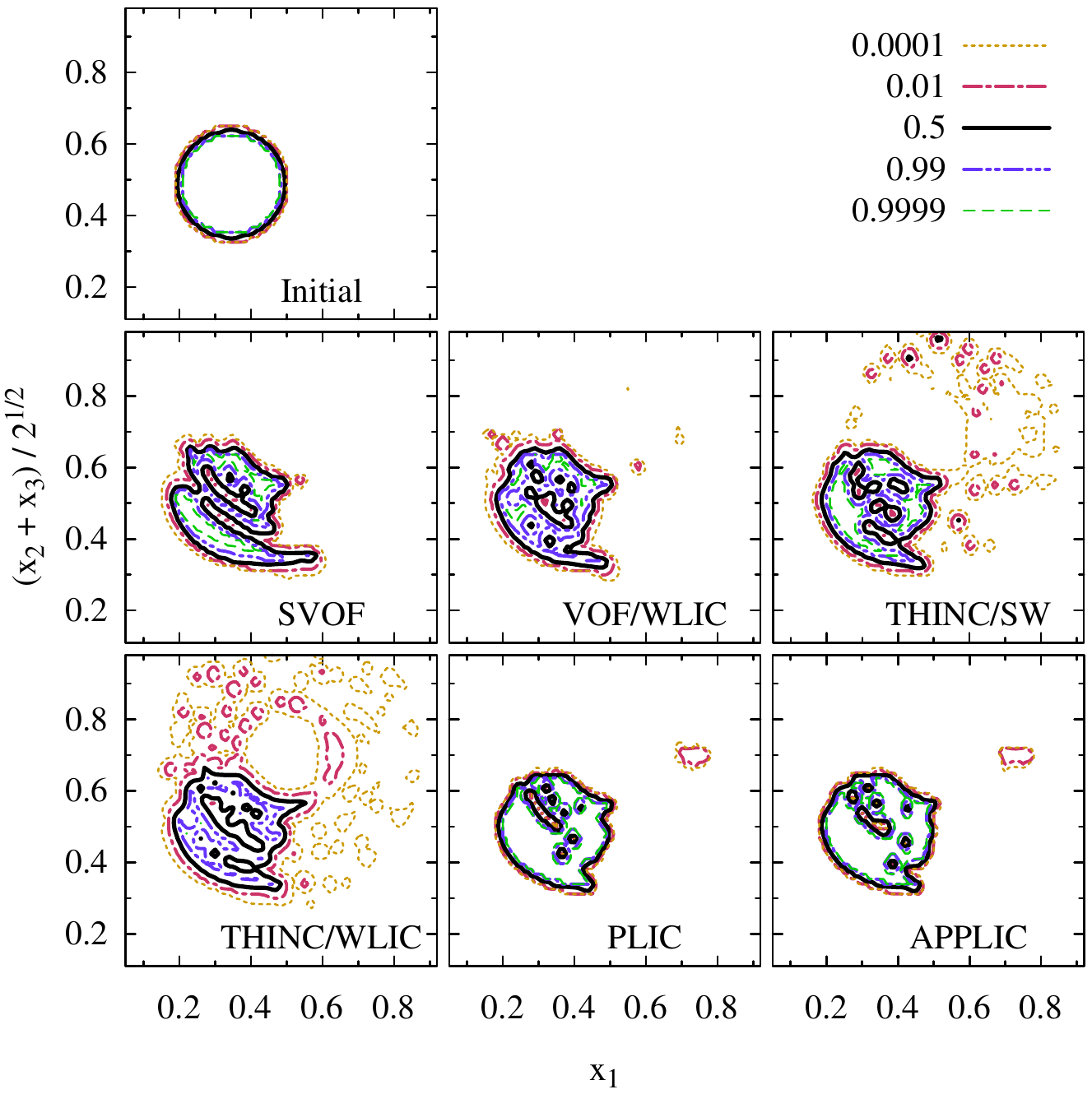}
\caption{\revised{Slice plots of the initial $(t=0)$ and final $(t=T)$ volume fractions
for test 3 at $x_2=x_3$,
using a $100\times100\times100$ grid with $\Delta t=0.0025$.
Volume fractions are drawn by contour lines of 0.00001, 0.01, 0.5, 0.99, and 0.99999.}}
\label{fig_test3_cntr}
\end{figure}

Figure \ref{fig_test3} compares the initial (exact) shape and
the final shapes
using a $100\times 100 \times 100$ ($\Delta x = 0.01$) grid with $\Delta t = 0.0025$,
where the maximum CFL number is approximately 0.5.
The SVOF and the VOF/WLIC methods produces significantly deformed shapes.
\revised{
Sliced plots of volume fractions are shown in Fig.~\ref{fig_test3_cntr}.
Artificial Low-density particles and voids are generated by all the VOF methods.
In particular the THINC/SW and the THINC/WLIC methods
generate many particles.
the PLIC and the APPLIC methods keep the interface widths more compact
than the SVOF, VOF/WLIC, THINC/SW, and THINC/WLIC methods.
}
The shapes produced by the THINC/WLIC method
have more prominent bumps than those by the PLIC and the APPLIC methods.
Table \ref{tbl_test3_1}  presents the $L_1$ errors and the corresponding convergence rates
for
$ 50\times  50\times  50$ $(\Delta x = 0.02)$,
$100\times 100\times 100$ $(\Delta x = 0.01)$, and
$200\times 200\times 200$ $(\Delta x = 0.005)$ grids
with $\Delta t= 0.005$, 0.0025, and 0.00125, respectively.
\revised{The errors of the PLIC and the APPLIC are the smallest
and the second smallest for all cases.}
\begin{table}[H]
\centering
\caption{
$L_1$ errors and convergence rates
using different grid sizes
for test 3.
}
\label{tbl_test3_1}
\begin{tabular}{cccccc}
\hline
Method
 & $ 25 \times 25 \times 25 $  & Rate  & $ 50 \times 50 \times 50 $  & Rate  & $ 100 \times 100 \times 100 $ \\
\hline
SVOF  & $1.23 \times 10^{-2}$& 0.72 & $7.47 \times 10^{-3}$& 1.21 & $3.23 \times 10^{-3}$ \\
VOF/WLIC  & $1.21 \times 10^{-2}$& 0.69 & $7.49 \times 10^{-3}$& 1.30 & $3.05 \times 10^{-3}$ \\
THINC/SW  & $1.13 \times 10^{-2}$& 1.10 & $5.24 \times 10^{-3}$& 1.69 & $1.62 \times 10^{-3}$ \\
THINC/WLIC  & $1.22 \times 10^{-2}$& 0.88 & $6.62 \times 10^{-3}$& 1.64 & $2.12 \times 10^{-3}$ \\
PLIC  & $1.07 \times 10^{-2}$& 1.10 & $4.98 \times 10^{-3}$& 2.04 & $1.21 \times 10^{-3}$ \\
APPLIC  & $1.05 \times 10^{-2}$& 1.04 & $5.09 \times 10^{-3}$& 2.01 & $1.26 \times 10^{-3}$\\
\hline
\end{tabular}
\end{table}

\subsection{Computational efficiency}
We investigate the computational efficiency of
the following method:
VOF/WLIC, THIC/WLIC, PLIC, and APPLIC.
The computational efficiency of the THINC/SW and the SVOF methods
is comparable to
that of the THINC/WLIC and the VOF/WLIC methods, respectively.
The computational time required to
evaluate fluxes for the 10,000,000 sample points in the set $S$,
defined in section \ref{def:setS},
for each method is used as a measure of efficiency.

The benchmark program is written in the C language.
Calculations using both single-precision and double-precision floating point arithmetic 
were conducted.
Three different computing platforms, summarized in Table \ref{tbl_systems},
were used to measure computational times.
\begin{table}[H]
\centering
\caption{Computing platforms to measure the computational time.}
\label{tbl_systems}
\begin{tabular}{cccc}
\hline
Platform & Computational processor & Compiler & Optimization options \\
\hline
I  & Intel Xeon E5-2643 v3$^a$          & Intel C Compiler 16.0 & {\tt -O3 -xCORE-AVX2}\\
II & Vector processor$^a$ of NEC SX-ACE & C++/SX 1.0            & {\tt -pvctl,noverrchk}\\
III& NVIDIA Tesla K40                   &  nvcc in CUDA 7.5     & {\tt -arch=sm\_35}\\ 
\hline
\multicolumn{4}{p{37em}}%
{\footnotesize $\!\!\!^{\rm a}$Only one core was used.}
\end{tabular}
\end{table}

The measured computational times are shown in Table~\ref{tbl_single}
(single-precision) and in Table~\ref{tbl_double}
(double-precision).
As shown in the tables,
the APPLIC method was 1.6-2.4 times faster than the PLIC method.
The VOF/WLIC and the PLIC methods were
the fastest and the slowest among the methods, respectively.
Although the algorithm of the THINC/WLIC method is
more simple than that of the APPLIC method,
the computational times of the two methods were comparable.
This is due to the calculations of transcendental functions,
which are much more time-consuming than the basic arithmetic operations.
To evaluate each flux,
the THINC/WLIC and the APPLIC methods requires five
(exp, log$\times 2$, and cosh$\times 2$) and four
(exp$\times 2$ and log$\times 2$) 
transcendental functions, respectively.
\begin{table}[H]
\centering
\caption{Computational times using single-precision floating point arithmetic.}
\label{tbl_single}
\begin{tabular}{cccc}
\hline
&\multicolumn{3}{c}{Computational times (ms)}\\
\cline{2-4}
Method    & Platform I & Platform II & Platform III\\
\hline
VOF/WLIC  & ~12.6    & ~5.66     & ~1.31\\
THINC/WLIC& ~78.4    & ~57.5     & ~1.74\\
PLIC      & 178.3    & ~97.8     & ~3.79\\
APPLIC    & ~75.4    & ~62.0     & ~1.66\\
\hline
\end{tabular}
\end{table}
\begin{table}[H]
\centering
\caption{Computational times using double-precision floating point arithmetic.}
\label{tbl_double}
\begin{tabular}{cccc}
\hline
&\multicolumn{3}{c}{Computational time (ms)}\\
\cline{2-4}
Method    & Platform I & Platform II & Platform III\\
\hline
VOF/WLIC  & ~25.7 & ~7.07 & ~2.67\\
THINC/WLIC& 241.5 & ~62.9 & ~4.04\\
PLIC      & 440.3 & 117.0 & ~7.55\\
APPLIC    & 195.6 & ~75.2 & ~4.15\\
\hline
\end{tabular}
\end{table}

\section{Conclusions}
We have presented a new PLIC-type VOF method called the APPLIC method.
In this method,
the complicated forward and inverse problems
that arise with the PLIC method
are approximately solved through
the use of the extremely simple formulae.
Accordingly, the APPLIC method is easier to develop and
to maintain the computational codes than the standard PLIC method.
The APPLIC method satisfies Eqs.~\eqref{eq:sgnprop}-\eqref{eq:f_range_loose},
which are essential to be satisfied for any VOF methods.


We conducted computational tests to compare accuracy of the APPLIC method
with other VOF methods; SVOF, VOF/WLIC, THINC/SW, THINC/WLIC, and PLIC.
The results of the tests show that the APPLIC method is
as accurate as the PLIC method
and more accurate than the SVOF, VOF/WLIC, THINC/SW, and THINC/WLIC methods.
It was demonstrated that
the computational time of the APPLIC method is shorter than that of the PLIC method
and comparable to that of the THINC/WLIC method.

\section*{Acknowledgments}
This research is partially supported by the Center of Innovation Program from Japan
Science and Technology Agency, JST.
The author thanks Akira Sou, Ippei Oshima, and Kensuke Yokoi
for engaging in insightful discussions and making useful comments.
\appendix
\section{}
In this appendix, we explain the derivation of Eq.~\eqref{eq_A_set}
to evaluate $C_\rA$.

The PLIC method assumes the shape of the dark fluid in the donor cell $\Omega$ as
the intersection of $\Omega$ and an oriented plane $\{\Bold{x} |\, \Bold{n}\cdot \Bold{x}<\alpha\}$.
The SZ algorithms
work with a unit cube
and a normal vector $\Bold{m}$
so that $\Bold{m} \ge 0$ and $\|\Bold{m}\|_1=1$.
Therefore, we apply a coordinate transformation
from $\Bold{x}$ to
$\Bold{x}'$
so that the donor cell $\Omega$ is mapped to
the unit cube $\{\Bold{x}'\in[0,1]^3\}$,
and the normal vector $\Bold{n}$ is mapped to a vector $\Bold{m}'$
satisfying the relation $\Bold{m}' \ge 0$.
The transformation, represented by $T_0$, is written as
\begin{align}
x'_l &= \frac{\sgn n_l}{\Delta x}(x_l - \xi_l),
\end{align}
where $\Bold{\xi}$ is the origin of the new coordinate,
which is chosen from
the eight vertices of $\Omega$
depending on the signs of $n_1$, $n_2$, and $n_3$
so that the image of the cell $\Omega$ coincides with the unit cube $\{\Bold{x}'\in[0,1]^3\}$.
The vertex $\Bold{\xi}$ is placed on the face $\phiOp$
if the signs of $m_I$ and $u_I$ are identical,
or on the face $\phi$ otherwise.
We can express the image of the oriented plane
$\{\Bold{x} |\, \Bold{n}\cdot \Bold{x}<\alpha\}$
under $T_0$ as
$\{\Bold{x}' |\, \Bold{m}'\cdot \Bold{x}'<\alpha'\}$,
where $\alpha'$ is the transformed plane constant.
The vector $\Bold{m'}$ is given by
\begin{align}
m'_l = \dfrac{\dfrac{n_l \Delta x}{\sgn n_l}}
 {
 \Big|\dfrac{n_1 \Delta x}{\sgn n_1}\Big|+
 \Big|\dfrac{n_2 \Delta x}{\sgn n_2}\Big|+
 \Big|\dfrac{n_3 \Delta x}{\sgn n_3}\Big|}
=\frac{\abs{n_l}}{\|\Bold{n}\|_1}.
\end{align}
The plane constant $\alpha'$ is determined by
the solution of the inverse problem 
\begin{align}
\alpha'=\alpha(C,\Bold{m}').
\end{align}

Let $\Omega'_\rA$ be the image of $\Omega_\rA$ under $T_0$.
We apply a further coordinate transformation $T_\rA$
from $\Bold{x}'$ to $\Bold{x}''$ so that $\Omega'_\rA$ is mapped to
the unit cube $\{\Bold{x}''\in[0,1]^3\}$
and all the components of the transformed normal vector
are nonnegative.
The transformation $T_\rA$ is written as
\begin{align}
x''_l =
\begin{cases}
x'_l               & \text{if $l\ne I$},\\
{x'_l}/{\absg}         & \text{if $l=I$ and $\Bold{\xi}$ is placed on $\phi$},\\
[{x'_l-(1-\absg)}]/{\absg} & \text{if $l=I$ and $\Bold{\xi}$ is placed on $\phiOp$}.
\end{cases}
\end{align}
We can express
the image of the oriented plane
$\{\Bold{x}' |\, \Bold{m}'\cdot \Bold{x}'<\alpha'\}$
under $T_\rA$ as
$\{\Bold{x}'' |\, \Bold{m}''_\rA\cdot \Bold{x}''<\alpha''_\rA\}$.
The normal vector $\Bold{m}''_\rA$ and
plane constant $\alpha''_\rA$ are given by
\begin{align}
m''_{\rA,l} &=
\begin{cases}
Q'_\rA m'_l\absg & \text{if $l=I$},\\
Q'_\rA m'_l & \text{if $l\ne I$},
\end{cases}
\\
\alpha''_\rA&=
\begin{cases}
Q'_\rA[\alpha'-m'_I (1-\absg)]&\text{if $\xi$ is placed on $\phi^*$},\\
Q'_\rA \alpha' &\text{if $\xi$ is placed on $\phi$},
\end{cases}
\end{align}
where $Q'_\rA$ is the normalization factor determined by
\begin{align}
Q'_\rA&=\frac{1}{m'_1 + m'_2 + m'_3 - m'_I + m'_I\absg}\notag\\
&=
\frac{1}{1 - m'_I(1-\absg)}.
\end{align}

Let $\Omega''_\rA$ be the image of $\Omega'_\rA$
under $T_\rA$.
The volume of the shape
$\Omega''_\rA\cap \{\Bold{x}'' |\, \Bold{m}''\cdot \Bold{x}''<\alpha''\}$
in the $\Bold{x}''$ coordinate system
is determined by solving the forward problem
$V(\alpha'',\Bold{m}'')$.
The partial volume fraction $C_\rA$ is thus obtained via Eq.~\eqref{eq_A_set}.

\section{}
In this appendix, we derive
the choice criterion $|\Til{\alpha}''_\rA - 1/2|> |\Til{\alpha}''_\rB - 1/2|$.

We define the approximation errors in $\Til{F}_\rA$ and $\Til{F}_\rB$
as
\begin{align}
\varepsilon_\rA &=(\Til{F}_\rA - F_{\rm PLIC})\sgn g,\\
\varepsilon_\rB &=(\Til{F}_\rB - F_{\rm PLIC})\sgn g.
\end{align}
Because $F_{\rm PLIC}\sgn\g = V(\alpha''_\rA, \Bold{m}''_\rA)\absg
=[C-(1-\absg)V(\alpha''_\rB, \Bold{m}''_\rB)],$
we have
\begin{align}
\varepsilon_\rA
&=[\Til{V}(\Til\alpha''_\rA,\Bold{m}''_\rA)
-
V(\alpha''_\rA,\Bold{m}''_\rA)]
\absg,\label{def:eps_A}\\
\varepsilon_\rB
&=
-[\Til{V}(\Til\alpha''_\rB,\Bold{m}''_\rB)
-
V(\alpha''_\rB,\Bold{m}''_\rB)]
(1-\absg),
\end{align}
where
The aim of this appendix is to find a quick and easy criterion
to guess whether $\abs{\varepsilon_\rA} < \abs{\varepsilon_\rB}$ or not.

The approximation error in $\Til V$ and $\Til \alpha$ are defined as
\begin{align}
\varepsilon_{\alpha}(V,\Bold{m}) &=
\Til\alpha(V,\Bold{m})
-\alpha(V,\Bold{m}),\\
\varepsilon_V(\alpha,\Bold{m}) &=
\Til V(\alpha,\Bold{m})
-V(\alpha,\Bold{m}).
\end{align}
Using Eqs.~\eqref{eq_A_app} and \eqref{eq_A_ap}, we obtain
\begin{align}
\alpha'&=\Til{\alpha}'
-\varepsilon_\alpha(C,\Bold{m}'),\\
\alpha''_\rA
&=\Til{\alpha}''_\rA
-Q'_\rA\varepsilon_\alpha(C,\Bold{m}').
\end{align}
The approximation error $\varepsilon_\rA$ is then given by
\begin{align}
\varepsilon_\rA
&=
\{\Til{V}(\Til\alpha''_\rA ,\Bold{m}''_\rA)
-
V[\Til{\alpha}''_\rA
-Q'_\rA\varepsilon_\alpha(C,\Bold{m}'),\Bold{m}''_\rA]\}
\absg\notag\\
&=
\left[
\varepsilon_V(\Til\alpha''_\rA, \Bold{m}''_\rA)
+
Q'_\rA
\varepsilon_\alpha(C,\Bold{m}')
\frac{\partial}{\partial\alpha}
{V}(\Til\alpha''_\rA,\Bold{m}''_\rA)
\right]
\absg+O(Q'_\rA\absg\varepsilon_\alpha^2).\label{eq:b7}
\end{align}
The value of $Q'_\rA\absg$ is included in the range $[0,1]$
because $m'_{I}\in[0,1]$ and  $\absg\in(0,1]$.


Let $V^*$ be an arbitrary value within $[0,1]$.
The following relation holds:
\begin{align}
\Til{V}[\Til{\alpha}(V^*)]=V[{\alpha}(V^*)]=V^*,\label{eq:b8}
\end{align}
Here the second arguments regarding $\Bold m$ are omitted for brevity.
On the other side,
the following can be obtained:
\begin{align}
{V}[{\alpha}(V^*)]
&=
{V}[\Til\alpha^* - \varepsilon_{\alpha}(V^*)]\notag\\
&={V}(\Til\alpha^*)-\varepsilon_{\alpha}(V^*)
\frac{\partial}{\partial \alpha}{V}(\Til\alpha^*)+O(\varepsilon_\alpha^2),
\label{eq:b9}
\end{align}
where $\Til\alpha^*$ stands for $\Til\alpha(V^*)$.
Using Eqs.~\eqref{eq:b8} and \eqref{eq:b9}, we have the relation between
$\varepsilon_V$ and $\varepsilon_\alpha$
as
\begin{align}
\varepsilon_V(\Til\alpha^*)=
-\varepsilon_{\alpha}(V^*)
\frac{\partial}{\partial \alpha}
{V}(\Til\alpha^*)
+O(\varepsilon_{\alpha}^2).\label{eq:b10}
\end{align}
Substituting Eq.~\eqref{eq:b10} into \eqref{eq:b7}
and neglecting the term $O(\varepsilon_\alpha^2)$,
we obtain
\begin{align}
\varepsilon_\rA
&\approx
\absg\delta_\rA
\frac{\partial}{\partial \alpha}
{V}(\Til\alpha''_\rA, \Bold{m}''_\rA),
\end{align}
with
\begin{align}
\delta_\rA
&=
Q'_\rA
\varepsilon_\alpha(C,\Bold{m}')
-\varepsilon_{\alpha}(\Til V''_\rA,\Bold{m}''_\rA).
\end{align}
Similarly,
\begin{align}
\varepsilon_\rB
&\approx
-
(1-\g)\delta_\rB
\frac{\partial}{\partial \alpha}
{V}(\Til\alpha''_\rB, \Bold{m}''_\rB),
\end{align}
with
\begin{align}
\delta_\rB
&=
Q'_\rB
\varepsilon_\alpha(C,\Bold{m}')
-\varepsilon_{\alpha}(\Til V''_\rB,\Bold{m}''_\rB).
\end{align}
Here $\Til V''_\rA$ and  $\Til V''_\rB$ stand for $\Til V(\Til\alpha''_\rA, \Bold{m}''_\rA)$
and $\Til V(\Til\alpha''_\rB, \Bold{m}''_\rB)$, respectively.

\begin{figure}[H]
\centering
\includegraphics[width=9cm]{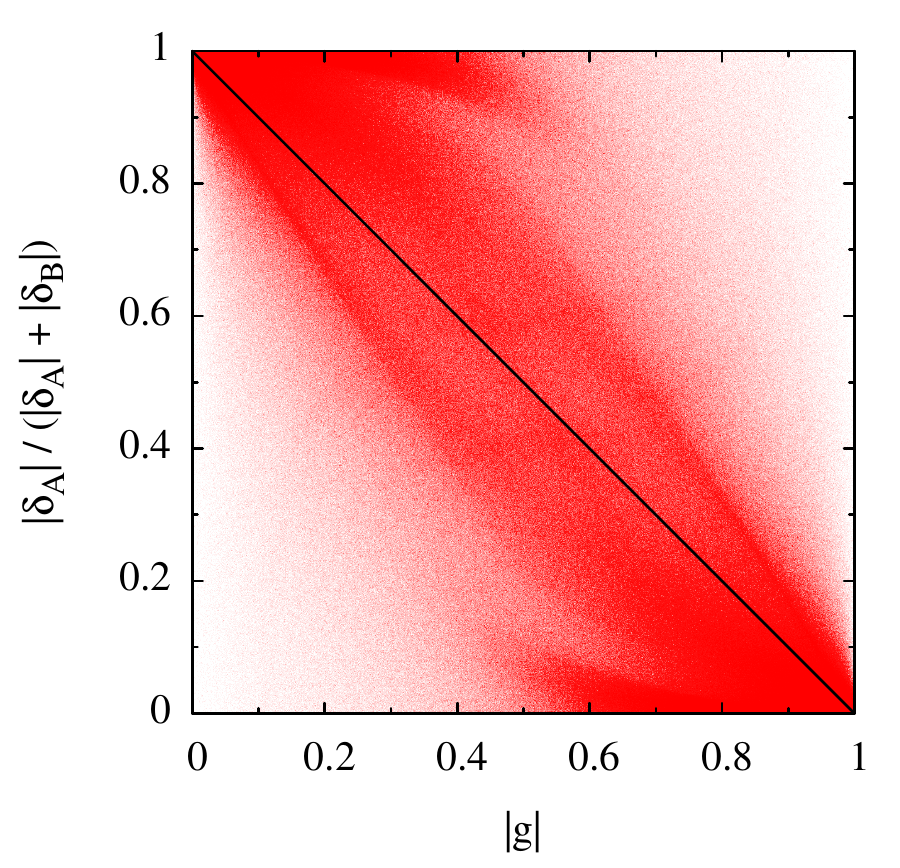}
\caption{Plot of 
$\absg$ versus $\abs{\delta_\rA}/(\abs{\delta_\rA}+\abs{\delta_\rB})$
for the sample points in the set $S$.
In this plot, a small red dot is drawn for each point.
The black solid line represents the line
$\abs{\delta_\rA}/(\abs{\delta_\rA}+\abs{\delta_\rB})=1-\absg$.}
\label{fig_blue_red}
\end{figure}

The value of $\delta_\rA$ becomes zero when $\absg$ is equal to one
(because $\Til V''_\rA=C$, $\Bold{m}''_\rA=\Bold{m}'$, and $Q'_\rA=1$ if $\absg = 1$).
Similarly, $\delta_\rB$ becomes zero when $\absg = 0$.
%
%
%
%
It is evident that
$\delta_\rA$ and $\delta_\rB$ are smooth functions with respect to $\absg$.
Considering these conditions,
we adopt the following approximations:
\begin{align}
\frac{\abs{\delta_\rA}}{
\abs{\delta_\rA}+\abs{\delta_\rB}}
&\approx 1-\absg,\label{eq:approx_delta_A}\\
\frac{\abs{\delta_\rB}}{
\abs{\delta_\rA}+\abs{\delta_\rB}}
&\approx \absg,\label{eq:approx_delta_B}
\end{align}
where we attach importance on simplicity rather than on accuracy.
Figure~\ref{fig_blue_red} demonstrates the accuracy of the approximations.
The data points are distributed roughly around the line $\abs{\delta_\rA}/({
\abs{\delta_\rA}+\abs{\delta_\rB}})=1  - \absg$.

By use of Eqs.~\eqref{eq:approx_delta_A} and \eqref{eq:approx_delta_B},
the inequality $\abs{\varepsilon_\rA} < \abs{\varepsilon_\rB}$ can
be written as
\begin{align}
\frac{\partial}{\partial\alpha}
{V}(\Til\alpha''_\rA,\Bold{m}''_\rA)
<
\frac{\partial}{\partial\alpha}
{V}(\Til\alpha''_\rB,\Bold{m}''_\rB),
\label{ineq:b17}
\end{align}
We employ the following approximation for simplicity:
\begin{align}
\frac{\partial}{\partial\alpha}
{V}(\alpha,\Bold{m}''_\rA)
=
\frac{\partial}{\partial\alpha}
{V}(\alpha,\Bold{m}''_\rB),
\end{align}
where $\alpha$ is an arbitrary value.
%
Because ${\partial V}/{\partial \alpha}$
is a monotonically nondecreasing function of $\alpha<1/2$
and has even symmetry to $\alpha=1/2$,
inequality~\eqref{ineq:b17} becomes
\begin{align}
\abs{\Til\alpha''_\rA - 1/2} > \abs{\Til\alpha''_\rB - 1/2}.\label{criterion}
\end{align}
Thus we obtain the criterion.

\clearpage
\bibliographystyle{elsarticle-num}
\bibliography{extre}

\begin{thebibliography}{10}
\expandafter\ifx\csname url\endcsname\relax
  \def\url#1{\texttt{#1}}\fi
\expandafter\ifx\csname urlprefix\endcsname\relax\def\urlprefix{URL }\fi
\expandafter\ifx\csname href\endcsname\relax
  \def\href#1#2{#2} \def\path#1{#1}\fi

\bibitem{hirt1981volume}
C.~W. Hirt, B.~D. Nichols, Volume of fluid ({VOF}) method for the dynamics of
  free boundaries, Journal of Computational Physics 39 (1981) 201--225.

\bibitem{rudman1997volume}
M.~Rudman, Volume-tracking methods for interfacial flow calculations,
  International Journal for Numerical Methods in Fluids 24 (1997) 671--691.

\bibitem{rider1998reconstructing}
W.~J. Rider, D.~B. Kothe, Reconstructing volume tracking, Journal of
  Computational Physics 141 (1998) 112--152.

\bibitem{scardovelli1999direct}
R.~Scardovelli, S.~Zaleski, Direct numerical simulation of free-surface and
  interfacial flow, Annual Review of Fluid Mechanics 31 (1999) 567--603.

\bibitem{pilliod2004second}
J.~E. Pilliod~Jr, E.~G. Puckett, Second-order accurate volume-of-fluid
  algorithms for tracking material interfaces, Journal of Computational Physics
  199 (2004) 465--502.

\bibitem{youngs1982}
D.~L. Youngs, Time-dependent multi-material flow with large fluid distortion,
  in: K.~W. Morton, M.~J. Baines (Eds.), Numerical Methods for Fluid Dynamics,
  Academic Press, New York, 1982, pp. 273--285.

\bibitem{li1995}
J.~Li, Calcul d'interface affine par morceaux (piecewise linear interface
  calculation), Comptes Rendus des Seances del' Academie des Sciences Paris,
  S\'erie IIb 320 (1995) 391--396.

\bibitem{noh1976}
W.~F. Noh, P.~Woodward, Slic (simple line interface method), Lecture Notes in
  Physics 24 (1976) 330--340.

\bibitem{yokoi2007efficient}
K.~Yokoi, Efficient implementation of {THINC} scheme: A simple and practical
  smoothed {VOF} algorithm, Journal of Computational Physics 226 (2007)
  1985--2002.

\bibitem{marek2008}
M.~Marek, W.~Aniszewski, A.~Bogus{\l}awski, Simplified volume of fluid method
  ({SVOF}) for two-phase flows, TASK Quarterly 12 (2008) 255--265.

\bibitem{xiao2005simple}
F.~Xiao, Y.~Honma, T.~Kono, A simple algebraic interface capturing scheme using
  hyperbolic tangent function, International Journal for Numerical Methods in
  Fluids 48 (2005) 1023--1040.

\bibitem{xiao2011revisit}
F.~Xiao, S.~Ii, C.~Chen, Revisit to the {THINC} scheme: a simple algebraic vof
  algorithm, Journal of Computational Physics 230 (2011) 7086--7092.

\bibitem{ii2012interface}
S.~Ii, K.~Sugiyama, S.~Takeuchi, S.~Takagi, Y.~Matsumoto, F.~Xiao, An interface
  capturing method with a continuous function: The {THINC} method with
  multi-dimensional reconstruction, Journal of Computational Physics 231 (2012)
  2328--2358.

\bibitem{scardovelli2000analytical}
R.~Scardovelli, S.~Zaleski, Analytical relations connecting linear interfaces
  and volume fractions in rectangular grids, Journal of Computational Physics
  164 (2000) 228--237.

\bibitem{hennessy2006computer}
J.~Hennessy, D.~Patterson, Computer Architecture: A Quantitative Approach, The
  Morgan Kaufmann Series in Computer Architecture and Design, Elsevier Science,
  2006.

\bibitem{scardovelli2003interface}
R.~Scardovelli, S.~Zaleski, Interface reconstruction with least-square fit and
  split {Eulerian--Lagrangian} advection, International Journal for Numerical
  Methods in Fluids 41~(3) (2003) 251--274.

\bibitem{aulisa2007interface}
E.~Aulisa, S.~Manservisi, R.~Scardovelli, S.~Zaleski, Interface reconstruction
  with least-squares fit and split advection in three-dimensional {Cartesian}
  geometry, Journal of Computational Physics 225~(2) (2007) 2301--2319.

\bibitem{weymouth2010conservative}
G.~Weymouth, D.~K.-P. Yue, Conservative {Volume-of-Fluid} method for
  free-surface simulations on {Cartesian-grids}, Journal of Computational
  Physics 229~(8) (2010) 2853--2865.

\bibitem{Wu2013739}
C.~S. Wu, D.~L. Young, H.~C. Wu, Simulations of multidimensional interfacial
  flows by an improved volume-of-fluid method, International Journal of Heat
  and Mass Transfer 60 (2013) 739--755.

\bibitem{vignesh2013noniterative}
T.~Vignesh, S.~Bakshi, Noniterative interface reconstruction algorithms for
  volume of fluid method, International Journal for Numerical Methods in Fluids
  73~(1) (2013) 1--18.

\bibitem{parker1992two}
B.~Parker, D.~Youngs, Two and Three Dimensional {Eulerian} Simulation of Fluid
  Flow with Material Interfaces, Atomic Weapons Establishment, 1992.

\bibitem{enright2002hybrid}
D.~Enright, R.~Fedkiw, J.~Ferziger, I.~Mitchell, A hybrid particle level set
  method for improved interface capturing, Journal of Computational Physics
  183~(1) (2002) 83--116.

\bibitem{zalesak1979fully}
S.~T. Zalesak, Fully multidimensional flux-corrected transport algorithms for
  fluids, Journal of Computational Physics 31~(3) (1979) 335--362.

\bibitem{leveque1996high}
R.~J. LeVeque, High-resolution conservative algorithms for advection in
  incompressible flow, SIAM Journal on Numerical Analysis 33~(2) (1996)
  627--665.

\end{thebibliography}

\end{document}